\documentclass[11pt]{article}
\usepackage{graphicx}
\usepackage{epstopdf}
\usepackage{amssymb}
\usepackage{amsmath}
\usepackage{hyperref}
\usepackage{float}
\usepackage{xcolor}
\usepackage[normalem]{ulem}

\usepackage{tabularx}
\usepackage{lineno} 
\usepackage{url}
\usepackage{setspace}
\providecommand{\keywords}[1]
{
  \small	
  \textbf{\textit{Keywords---}} #1
}

\def\beq{\begin{equation}}
\def\eeq{\end{equation}}
\def\bea{\begin{eqnarray}}
\def\ena{\end{eqnarray}}

\def\b{\beta}
\def\k{\kappa}
 
\def\eps{\epsilon}

\def\g{\gamma}
 \long\def\drop#1{}

\begin{document}
\doublespacing
\title{The effect of predation on the dynamics of Chronic Wasting Disease in deer}

\author{Cody E. FitzGerald\textsuperscript{1 } and James P. Keener\textsuperscript{2}\footnote{deceased} \\
 \textsuperscript{1} Department of Engineering Sciences and Applied Mathematics, Northwestern University,\\
Evanston, IL 60208, USA\\ and\\ \textsuperscript{2} Department of Mathematics, University of Utah,\\ Salt Lake City, Utah, 84112 USA.}



 \maketitle 
 

\keywords{Chronic Wasting Disease, population ecology, control, management, bifurcation theory, wolves, dynamical systems, disease ecology}
 
 \section{Abstract}
 
 Chronic Wasting Disease (CWD) is a neurological disease impacting deer, elk, moose, and other cervid populations and is caused by a misfolded protein known as a prion.  CWD is difficult to control due to the persistence of prions in the environment. Prions can remain infectious for more than a decade and have been found in soil as well as other environmental vectors, such as ticks and plants. Here, we provide a bifurcation analysis of a simple mathematical model of CWD spread in a cervid population and use a modification of the Gillespie algorithm to explore if predators can be used as an ecological control strategy to limit the spread of the disease in several relevant scenarios. We then use several analytical probabilistic arguments to show that lowering the susceptible population is also beneficial to controlling CWD outbreaks. Finally, we consider a more complex model of CWD spread in a cervid population in which predators are assumed to be dynamic. Here, we find that, again, predators may be used to control CWD outbreaks, assuming they selectively prey upon infected cervids at a sufficiently high rate. From our analysis, we conclude that wolves may be used as an effective control strategy to limit the spread of CWD in cervid populations, and hunting or other means of lowering the susceptible population are beneficial to controlling CWD. 
 
\section{Introduction}


Chronic Wasting Disease (CWD) is a prion-mediated neurodegenerative disease that has been observed in cervids, such as elk, deer, and moose, and was first described in 1980 \cite{williams1980chronic}. Prions are misfolded and infectious proteins that can cause neurological disease, such as CWD. CWD is transmitted both directly between susceptible animals and indirectly through environmental exposure. Cervids can become infected through contact with saliva, blood, and other bodily fluids of an infected animal or through indirect environmental infections from water sources, plants, soil, or potentially insects. In recent years, wild populations of cervids have exhibited high levels of CWD prevalence. The 2024 Chronic Wasting Disease Surveillance Report from the Wyoming Game and Fish Department reported that the prevalence of CWD in some populations of mule deer is as high as 66.3\% \footnote{The report can be found here \url{https://wgfd.wyo.gov/wyoming-wildlife/wildlife-disease-and-health/chronic-wasting-disease} using the drop-down Report menu at the bottom of the website.}. CWD has proven difficult to control or eradicate, due to the time scale associated with the persistence of CWD prions in the environment. Prions decay slowly and can remain latent and infectious for more than a decade \cite{GSP_2006,wiggins_2009}. Many mathematical models have been developed to help understand the disease spread in deer populations, evaluate the efficacy of control strategies, and infer the modes of transmission. We briefly review several of the published models, but our summary should not be considered exhaustive. 

One of the earlier discrete population models examined the spread of CWD and control strategies \cite{miller2000epizootiology,gross2001chronic}, but the modeling assumptions, in particular frequency-dependent transmission, were reevaluated in the follow-up paper \cite{schauber2003chronic}. Miller and co-authors developed and attempted parameter estimation of competing SIR-type models that describe transmission of CWD in a population of cervids through different mechanisms---though at least one of the models considered appears to exhibit a lack of structural identifiability, which we briefly comment on in the Discussion \cite{miller_2006}. They found support for indirect transmission of CWD using a model selection approach \cite{miller_2006}. A different study also suggests that the disease will spread via indirect transmission due to the persistence of prions in the environment \cite{almberg2011modeling}. The model presented in \cite{miller_2006} was further modified for further mathematical and control-based investigations. 

Sharp and Pastor modified the model presented in \cite{miller_2006} to include logistic growth in the susceptible population and performed a mathematical analysis of the system \cite{SharpPastor2011}. Wild and co-authors later extended the model presented in \cite{miller_2006} to include the effects of predation from wolves \cite{wild_miller_2011} and analyzed the system in a non-oscillatory parameter regime, using parameter values estimated in \cite{miller_2006} and other studies. An earlier report to the National Park Service by Hobbs also analyzed wolf-predation as a control strategy using a similar deterministic model in an oscillatory parameter regime \cite{hobbs2006model}. Barbera and Pollino performed a mathematical analysis of a PDE \cite{Barbera_2023} that was inspired by the model presented in \cite{SharpPastor2011}. Reyes and co-workers designed a spatiotemporal causal inference scheme to examine the effects of culling strategies using a PDE-based approach \cite{Reyes_2025}. Potapov and co-authors used a complex age-structure differential equation model to examine the influence of harvesting strategies on population dynamics and disease suppression \cite{potapov_2016}. Vasylieva and co-authors performed a mathematical analysis of a differential equation model that describes CWD spread in a population of deer, including both indirect and direct transmission \cite{Vasylieva_2015}. They also attempted a parameter estimation of the system, using upper and lower bounds for parameter estimation based on estimates of parameters found in \cite{miller_2006}. A different study examined the influence of culling strategies to control CWD spread using a differential equation model that incorporated seasonality into a population model of cervids \cite{Oraby_2014}.  Others created an integrodifference equation model that incorporated direct and indirect transmission and long-distance dispersal from juvenile deer to understand CWD spread in a population of white-tailed deer \cite{McClure_2024}. Several groups have examined harvest strategies using mathematical models that delineate the deer population by sex \cite{al2012modeling,Al-aryday_2016,rogers2022epidemiological}. Control investigations into CWD transmission in deer populations are summarized in \cite{uehlinger2016systematic}. 

Recently, agent-based models, statistical models, and machine learning approaches have been used to examine CWD transmission in deer populations. An agent-based model has been used to examine the spread of CWD in deer populations and possible management strategies to limit CWD \cite{thompson_2024}. A recent long-term statistical analysis used movement data collected from 596 deer to examine how seasonal behaviors, habitat selection, and home range size influence CWD transition in southwest Wisconsin \cite{Gilbertson_2025}. Machine learning approaches have also been applied to predicting CWD incidence at the county level \cite{ahmed2024predicting}. While machine learning approaches are often useful for accurate forecasting and prediction, they typically do not shed light on the underlying mechanisms that underpin the system and its dynamics, i.e., what drives CWD outbreaks and how such outbreaks may be controlled. Here, we attempt to study the control question using a simple, interpretable \textit{mechanistic} model. 

In this paper, we present and analyze a relatively simple model to understand the CWD spread in a population of deer and explore the role of predators, such as wolves, as an ecological control strategy. In our initial model, we treat the predator density as constant. Our model is inspired by previous mathematical models \cite{SharpPastor2011,wild_miller_2011,hobbs2006model}. We first perform a bifurcation analysis to demonstrate the effect of predation on the dynamics of the disease and then simulate the model stochastically using a modified form of the Gillespie algorithm \cite{gillespie1977}. We show that the presence of wolves in all cases lessens the severity of a CWD outbreak and can lead to the elimination or at least significant reduction of the severity of the disease in the deer population. This is true even if the control intervention is applied after CWD has spread in the population. The mechanism of this reduction is easy to understand, as predation increases the death rate of infected animals and thereby decreases the amount of spread of prions from the infected animals. We then make analytic probabilistic arguments to show that reducing the susceptible population, not just the infected population, is also beneficial to limit future outbreaks. Finally, we study a more complex, realistic model, in which the predator is assumed to be dynamic. In this case, we find that predators can still be an effective control strategy for CWD outbreaks, but this intervention may not be expected to fully eradicate the disease. Our analysis shows that CWD outbreaks in deer may be controlled or limited by a combination of predator-based control strategies, which lower the infected population, and hunting or other harvesting strategies to limit the size of the susceptible population. 

%
%
%
 

 \section{A Continuous Variable Population Model}
 \label{bif}
  The population dynamics model that we begin with was first presented and studied in \cite{SharpPastor2011}, and is given by
\bea
\frac{dS}{dt} &=& rS(1-\frac{S}{K})-\gamma SE,\label{eq:m1}\\
\frac{dI}{dt} &=& \gamma SE-\mu_i I,\label{eq:m2}\\
\frac{dE}{dt} &=& \epsilon I - \mu_e E.\label{eq:m3}
\ena

Here, $S$ is the density of susceptible and uninfected animals, $I$ is the density of infected animals, and $E$ is the density of prions in the environment (primarily in the soil).   The meaning of the terms of the model is that $S$, in the absence of prions, grows according to logistic dynamics, but becomes infected at a law of mass action rate from prions.  The infected animals have a death rate $\mu_i$. Prions are shed by infected animals at rate $\eps$, and degrade at rate $\mu_e$.
\drop{
\title {Parameter values for the CWD model}

\label{table:stability}
\begin{tabular}{|c|c|c|c|}
 \hline
 $r$&1.5 yr$^{-1}$
 & $\gamma$&0.8 yr$^{-1}$(prion~density)$^{-1}$\\
 \hline $\mu_i$&1.0 yr$^{-1} $
 & $\epsilon $&0.1  yr$^{-1}$(prion density)(deer density)$^{-1}$\\
 \hline   $K$&0-50(100 km$^{2})^{-1}$& $\mu_e$&0.1-0.2 yr$^{-1}$\\
  \hline
 \end{tabular}
 \end{center}
\end{table}
}

A slightly earlier model \cite{wild_miller_2011} included the effects of predation  using  the equations
\bea
\frac{dS}{dt} &=& r(S+I)(1-\frac{S+I}{K})-S(\gamma E+m)-\delta \rho_s(S,I)S,\\
\frac{dI}{dt} &=& \gamma SE-(\mu_i +m) I-\delta \rho_i(S,I)I,\\
\frac{dE}{dt} &=& \epsilon I - \mu_e E.\label{eq:3}
\ena
where $\rho_s(S,I)= {S+I\over S+vI}$, $\rho_i(S,I)=v(1-c){S+I\over S+vI}$.

 The analysis of the model in \cite{SharpPastor2011} showed that there are three regions of parameter space with distinct dynamic behaviors: a region with no infected animals or prions, a region in which the disease is endemic, but steady, and a region in which there are periodic (oscillatory) outbreaks of the disease, during which the animal population falls catastrophically low.  They did not study the effect of predation on the dynamics of the disease.  The model in \cite{wild_miller_2011} was used to study the effects of predation through the parameter, $\delta$, but failed to notice the existence of periodic solutions and the effect of predation on those oscillations. Separately, a rigorous mathematical analysis of a closely related model is given in \cite{maji2018deterministic}.

The model considered here uses a slightly different density dependence on birth and death and a simplified predation effect, motivated by simple predator-prey interactions,
  \bea
\frac{dS}{dt} &=& bS(1-\frac{S+ I}{K'})-dS-\gamma _eSE -\gamma_i SI-\rho_sSW,\label{eq:m7a}\nonumber\\
&=& rS(1-\frac{S+ I}{K})-\gamma_e SE -\gamma_i SI-\rho_sSW,\label{eq:m7}\\
\frac{dI}{dt} &=& \gamma_eSE+\gamma_i SI-\mu_i I-\rho_iIW,\label{eq:m8}\\
\frac{dE}{dt} &=& (\epsilon_i+\epsilon_d\mu_i)I   -\mu_e  E\equiv \epsilon I\ -\mu_e  E \label{eq:m9},
\ena
where $W$ is the density of the predator (wolf) population, assumed here to be constant, and $\rho_s$ and $\rho_i$  are both constant, quantifying the effect of predation.  Here  $b$ is the zero population birthrate, $K'$ is the population density at which the birthrate is zero, $d$ is the natural death rate of healthy animals, and $r=b-d$, $K=(1-{d\over b})K'$, are both assumed to be positive.   Further, the model incorporates the assumption that infected animals contribute to the carrying capacity, but not to the birth of susceptible animals. The disease is assumed to spread in one of two ways, through contact with prions in the environment (rate $\gamma_e$) and through direct contact between infected and susceptibles (rate $\gamma_i$). Prions are deposited into the environment because of shedding by infected animals (at rate $\epsilon_i I$) or death of infected animals (at rate $\epsilon_d\mu_i I$). Notice that the units for $\epsilon_i$ (yr$^{-1}$(prion density)(deer density)$^{-1}$) are different than those for $\epsilon_d$ ( (prion density)(deer density)$^{-1}$). Here, our goal was to develop and analyze a model that is biologically motivated, contains both indirect and direct transmission pathways, contains monomial terms rather than rational terms for ease of use in analytic bifurcation theory, is simple to understand, and is interpretable. Parameter values used for this study are shown in Table \ref{table:parameters}, although we briefly mention possible concerns about structural identifiability of models used in \cite{miller_2006} in Section \ref{Discussion}. Without loss of generality, $\gamma_e$ can be taken as one by rescaling the units on $E$.

\begin{table}[ht] 
  \centering 

    \begin{tabularx}{\textwidth}{|l|X|X|X|} 
    \hline
Parameter&Definition&Value&Reference\\
 \hline
 $b$& birth rate$^a$ &0.6 yr$^{-1}$ &\\
 &at population =0&&\\
  \hline  
   $d$& death rate$^b$ &0.1 yr$^{-1}$ & \\
    \hline  
   $r=b-d$& effective birthrate &0.5 yr$^{-1}$ & \\
  \hline $\gamma_i$&direct contact infection rate&0.1 yr$^{-1}$(infected~animal~density)$^{-1}$&\cite{miller_2006}\\
   \hline $\gamma_e$&prion transmission rate&1yr$^{-1}$(prion~density)$^{-1}$&wlog\\
 \hline $\mu_i$&CWD death rate &0.6 yr$^{-1} $
 &\cite{miller_2006}\\\hline  $\epsilon $&Rate of excretion&0.1  yr$^{-1}$(prion density)(deer density)$^{-1}$&\cite{miller_2006}\\
 & of infectious material&&\\
 \hline  $\mu_e$&Rate of decay&0.1-0.2 yr$^{-1}$&\cite{miller_2006} \cite{SharpPastor2011} \\
 & of infectious material &&\\
 \hline   $K$&Carrying capacity &0-100(100 km$^{2})^{-1}$&\\
  \hline   $K'$&=${K/ (1-{d\over b})}$ &0-100(100 km$^{2})^{-1}$&\\
  \hline

    \end{tabularx}
       \caption{a)   birth rate at population =0  (=1.8 fawns per female deer per year) b) death rate  = 0.1 (probability of survival per year = 0.9).}  
       \label{table:parameters}
    \end{table}

    It is likely that $\rho_s<\rho_i$, since wolves prey primarily on aged and disabled deer, and the percentage of such among the susceptible population is certainly smaller than among the CWD-infected deer. 
         
 There are three possible steady solutions for this model: the extinct state with $S=0$,  the disease-free state, and the disease endemic state.  The disease-free state has
  \beq S= K(  1-{\rho_s\over r}W ), \quad I=0,\quad E=0.
  \label{eq:disease_free}
\eeq
  
This solution requires that $\rho_sW<r$, i.e., that predation of the wolves on healthy animals is not too large.  

The disease endemic state is given by
\bea
  S&=&  {\mu_e\over \g}(\mu_i + \rho_iW),\label{eq:endemic_state}\\
I&=&{\mu_e\over \g}{r(K\gamma - \mu_e\mu_i)  - (K\gamma \rho_s  + \mu_er\rho_i )W\over  K\gamma + r\mu_e },\\
 E&=& {\eps \over \mu_e} I, 
\ena
where  $\gamma =  \eps\g_e + \g_i\mu_e$.
     Clearly, this solution only exists  
          provided $r(K\gamma - \mu_e\mu_i)  - (K\gamma \rho_s  + \mu_er\rho_i )W>0$.
   
   One can also show that the disease-free state is stable if $r(K\gamma - \mu_e\mu_i)  - (K\gamma \rho_s  + \mu_er\rho_i )W<0$, and unstable if not.  It is immediately apparent that the presence of predators decreases the likelihood of an endemic disease state existing and stabilizes the disease-free state.   Indeed, if $W$ is large enough, there is only the disease-free state.   Furthermore, in the disease-free state, the presence of predators decreases the $S$ population (see (\ref{eq:disease_free})), whereas in the endemic disease state, if it exists, the presence of predators increases the $S$ population, while it decreases the level of the infected population   $I$ as well as the environmental load of prions, $E$ (see (\ref{eq:endemic_state})).

  Using XPP, Sharp and Pastor \cite{SharpPastor2011} showed that there is a curve in parameter space at which there is a Hopf bifurcation for the equations (\ref{eq:m1}-\ref{eq:m3}).  It is a  direct calculation (i.e., find the location in parameter space at which the Jacobian of the  system (\ref{eq:m1}-\ref{eq:m3}) has purely imaginary eigenvalues \cite{keener_resultant}) to show that this curve is given    by 
\beq
H(\kappa)=\kappa^2 - \kappa(\mu_e^2 + 3 \mu_e\mu_i + \mu_i^2) - \mu_e\mu_ir (\mu_e+\mu_i)=0,
\eeq
where $\kappa = K \eps\g$.  Numerical simulations verify that there are stable periodic solutions for parameter values for which $H(\k) > 0$.  Since $H(\k) = 0$ has a unique positive root, this implies that there are periodic solutions for $\kappa$ sufficiently large.
 
We can also find the curve of Hopf bifurcation points for equations (\ref{eq:m7}-\ref{eq:m9}).    It is given by $H_W=0$, where 
\bea
H_W&=& K^3\eps\g_e \gamma^2(r - \rho_sW)(\gamma-M\g_i )\nonumber\\
&&-K^2\gamma r(\g_i\mu_e(2M^2\g_i\mu_e - 3\gamma M^2 - 3\gamma M\mu_e) +\b \gamma^2)\nonumber\\
&&-K\mu_er^2(\g_i\mu_e(M^2\g_i\mu_e - 3\gamma M^2 - 2\gamma M\mu_e) + \b \gamma^2)\nonumber\\
&&-Mr^3\mu_e^3(\gamma-M\g_i  ),\label{eq:hopf_curve}
\ena
where $\b =   M^2 + (3\mu_e +   r -  \rho_sW)M +  \mu_e^2$, $\gamma = \eps\g_e + \g_i\mu_e$,  $M=\mu_i +\rho_iW$.

\begin{figure}[ht]
 \center
 \includegraphics[height=5cm]{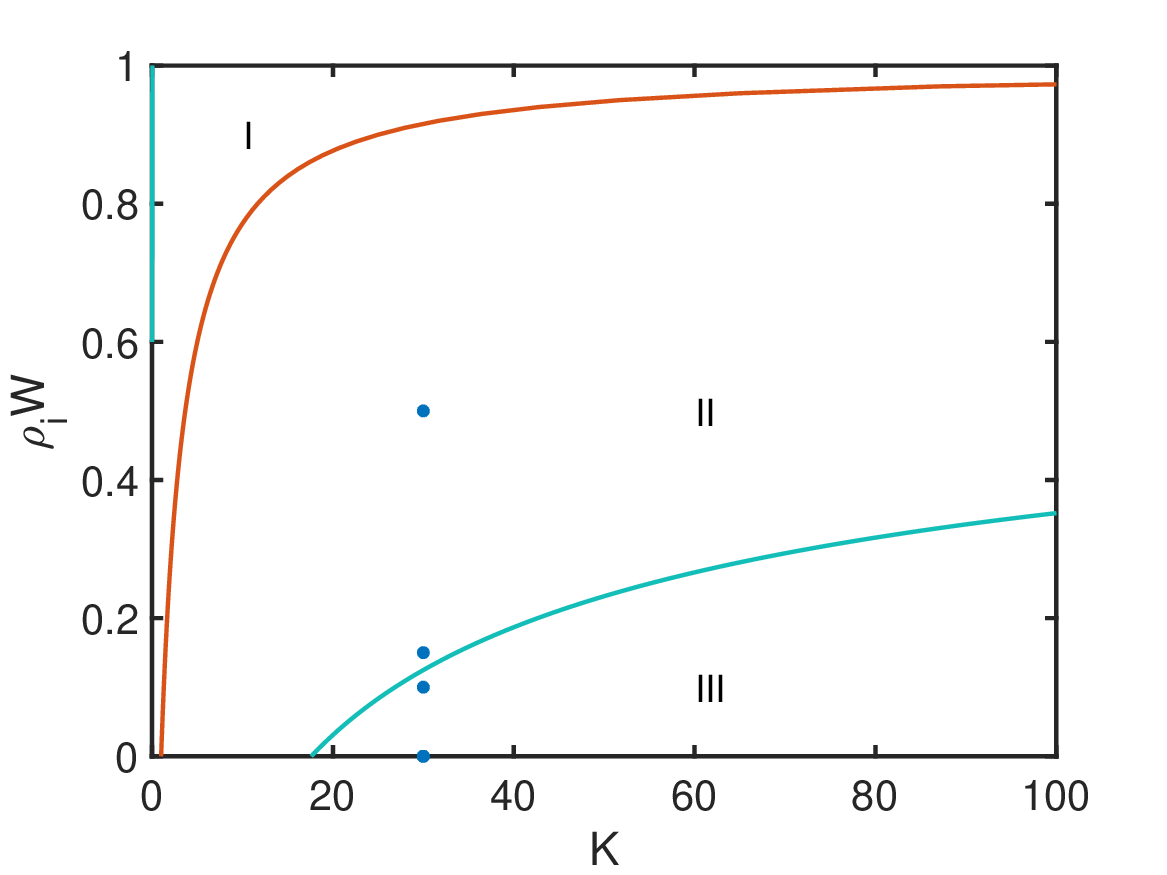}
  \caption{Regions for Chronic Wasting Disease: I: No disease, II: Disease is endemic, III: Disease has oscillatory outbreaks, IV: Extinction (not shown here),  where $\frac{\rho_s}{\rho_i} = 0.5$.  The curve separating region I from region IV is $\rho_sW=r$; The curve separating region I from region II is $r(\kappa    - \mu_i\mu_e)- W(\mu_er\rho_i +\kappa \rho_s) =0$, and the curve separating region II from region III is $H_W(\k) =0$ (Eq. \ref{eq:hopf_curve}).  Asterisks show locations of the solutions shown in Fig.~\ref{fig:3de} with $K=30$, and $\rho_iW=0.0,  0.1, 0.15, 0.5$, and other parameters as in Table \ref{table:parameters}.}  
 \label{fig:1de}
 \end{figure}
 
 These features of predation are depicted in Fig.~\ref{fig:1de}. It is shown that parameter space is divided into four regions: I, in which there is no disease; II, in which the disease is endemic; III, in which the disease is oscillatory; and IV, in which the deer population is extinct because of over-predation.  The boundaries between these regions are shown plotted in $\rho_iW-K$ parameter space, showing the stabilizing effect of predators.

 \begin{figure}[ht]
 \center
 \includegraphics[height=5cm]{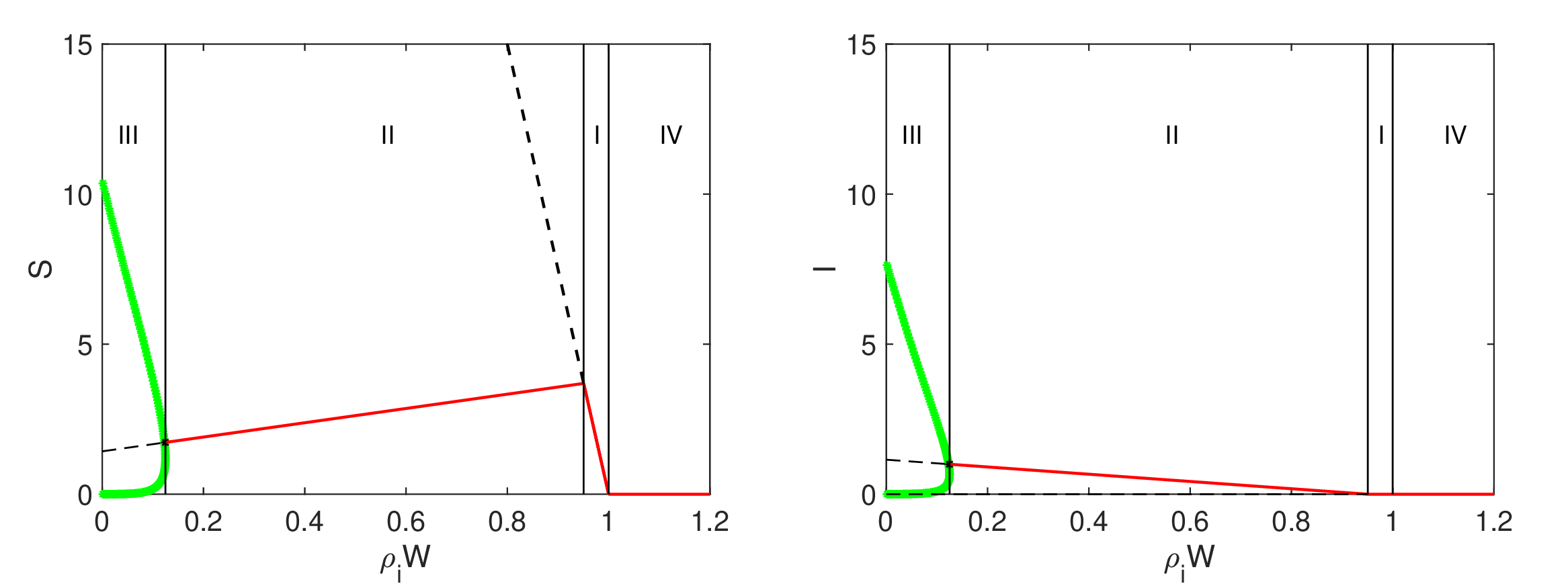}
  \caption{Bifurcation Diagram for Chronic Wasting Disease, corresponding to a vertical slice of Fig.~\ref{fig:1de}   at $K=30$: I: No disease, II: Disease is endemic, III: Disease has oscillatory outbreaks, IV: Predation leads to extinction.  Green curves denote the maximum and minimum of stable oscillations, red curves denote stable steady solutions, and black dashed curves denote unstable steady solutions.  For these plots,    $\frac{\rho_s}{\rho_i} = 0.5$, $\mu_e=0.2$, and other parameters as in Table \ref{table:parameters}.}
 \label{fig:2de}
 \end{figure}
 
  \begin{figure}[ht]
 \center
 \includegraphics[height=10cm]{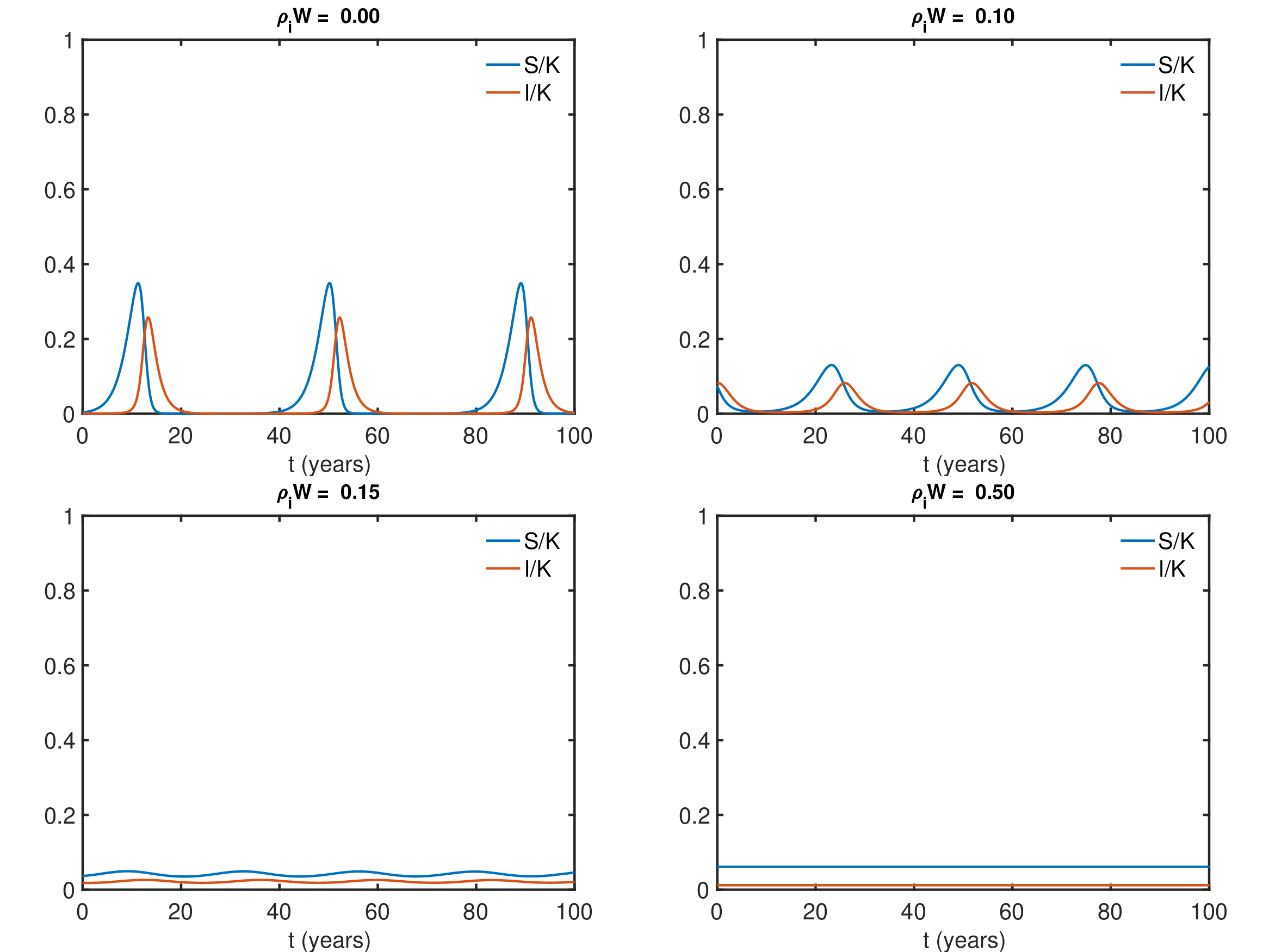}
  \caption{Solution of the differential equation system with (Upper Left) $\rho_iW=0$, (oscillatory outbreaks), (Upper Right) $\rho_iW=0.1$ (suppressed oscillations),  $\rho_iW=0.15$, and $\rho_iW=0.5$ (no oscillations, stable endemic state).  For these plots,    $\frac{\rho_s}{\rho_i} = 0.5$,  $K=30$, $\mu_e=0.2$, and other parameters as in Table \ref{table:parameters}.}
 \label{fig:3de}
  \end{figure}

Not only does the presence of a predator decrease the likelihood of disease, but it also reduces the prion load of the environment and decreases the amplitude, hence the severity, of oscillatory outbreaks when they occur.  This can be seen in numerical simulations, as demonstrated in Fig.~\ref{fig:3de}.  Specifically, in these plots are shown the numerical simulation of the ODE system for the fixed value of $K=30$ for different values of $\rho_iW = 0,   0.1, 0.15, 0.5$, placing these in regions II and III (see asterisks in Fig.~\ref{fig:1de}).  Noticeably, the effect of predation is to reduce the severity of the disease, at the cost of reducing the overall population size below its carrying capacity.  

\section{Stochastic simulation} 
 \label{stoch}

Because deer populations are discrete and not too large, a continuous variable model may be inappropriate, and there might be insight to be gained from stochastic simulations of this process.
To this end, we consider the following seven  reactions:
\bea
R_1: &&s\rightarrow s+1 \quad {\rm (Birth~of~deer)}\\
R_2: &&s\rightarrow s-1 \quad {\rm (Death~of~deer)}\\
R_3:&&s, i\rightarrow s-1,i+1 \quad {\rm (Infection~of~deer~ by ~E)}\\
R_4:&&s,  i\rightarrow s-1,i+1 \quad {\rm (Infection~of~deer~ by~}i)\\
R_5:&&i\rightarrow i-1\quad {\rm (Death~of~infected~deer)}\\
R_6:&&s \rightarrow  s-1 \quad {\rm (Predation ~of~uninfected~deer)}\\
R_7:&&i \rightarrow i-1  \quad {\rm (Predation ~of~infected~deer)}
\ena
where $s$ and $i$ represent the number of susceptibles and infected, respectively, at any given time. Since we are taking the wolf population to be constant, it is not necessary to include wolf dynamics here. We need to get the reaction rates correct for the stochastic simulation.  We start with some representative area $A$ that we want to consider in our simulation, and realize that the unit of densities is set by some area $A_0$, in this case, $A_0=100$km$^2$.  This means, for example, that if $K$ is the carrying capacity density, then $K$ is also the number size of the carrying capacity in an area of size $A_0$.  Therefore, the size of the carrying capacity for an area of size $A$  is $K{A\over A_0}$. Now, let $s$ and $i$ be the integers that the densities $S$ and  $I$ represent. Then, if $S$ is the density of susceptibles in the area $A_0$,   the number of susceptibles in the area $A$ is $s=S{A\over A_0}$, and similarly for infected animals, $i = I{A\over A_0}$.  

The reaction rates for the stochastic process are, therefore,
\bea
R_1:&& r_1 = b s(1-{A_0\over K' A}(s+i)),\\
R_2: &&r_2 =  d s,\\
R_3:&& r_3 = \gamma_e E s,\\
R_4:&& r_4 = \gamma_i {A_0\over A}si,\\
R_5:&& r_5 = \mu_i i,\\
R_6:&&r_6= s\rho_s  W,\\
R_7:&&r_7= i\rho_i W,
 \ena
 Here, $E$ remains a continuous, not discrete,  variable, and so is governed by the differential equation  $\frac{dE}{dt} =   \frac{\epsilon A_0}{A}i-\mu_e E$.  
 We would like to use these reaction rates with the Gillespie algorithm \cite{gillespie1977} to do stochastic simulations.  However,   since $E$ is continuously changing in time,  the Gillespie algorithm is not directly applicable.  Instead, we use ``Poisson thinning" to account for this fact, as follows\cite{Lewis_shedler_79,Ogata_81,Ross_1997}: Since between reactions, $E(t)$ is a monotone function, bounded between $E(t_0)$ and $\epsilon \frac{iA_0}{\mu_e A}$, where $t_0$ is the last reaction time, we let $E^*={\rm max}[E(t_0),\epsilon \frac{i  A_0}{\mu_e A}]$, and introduce a modified reaction $R_3^*$ (since $R_3$ depends on $E$), with reaction rate $r_3^* = \gamma s E^*$ (which is the fastest possible reaction rate), and then proceed with the usual Gillespie determination of next reaction time and next reaction.  However, whenever the next reaction is to be $R_3^*$ at time $t^*$,   the reaction is split into two, only one of which is implemented: a null reaction for which no change to $s$ or $i$ is made with probability $\frac{E^*-E(t^*)}{E^*}$, and the reaction that reduces $s$ by one and increases $i$ by one with probability $\frac{E(t^*)}{E^*}$.

  Because $K$ is in the range of 0-100(100 km$^2$)$^{-1}$, so $A_0 = 100$km$^2$, to have a reasonable number of deer, it must be that $A=1000$km$^2$, at least. This also needs to take into account the territory size of a wolf pack, which can vary widely from less than 100 km$^2$ to more than 4000 km$^2$ \cite{mech2019wolves}. In our stochastic simulations, we take $A=2000$km$^2$ unless otherwise noted. 

 \begin{figure}[ht]
 \center
 \includegraphics[height=5cm]{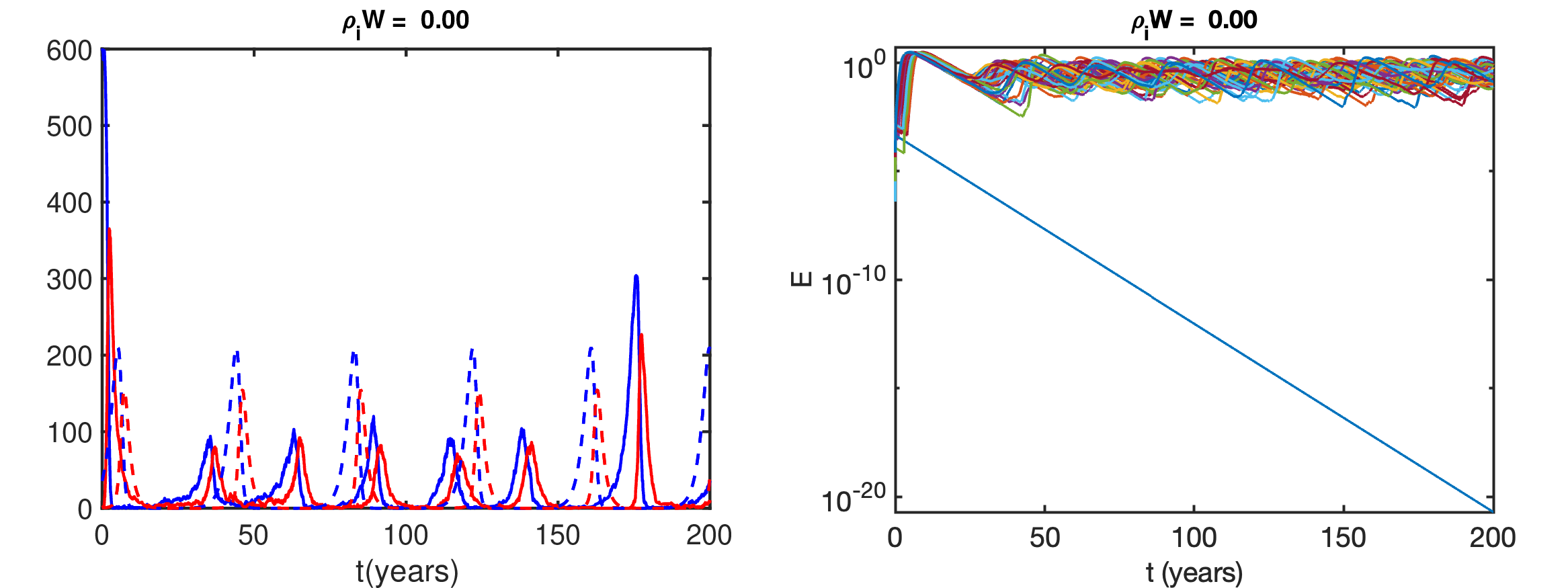}
  \caption{Example of stochastic simulation showing (Left) $s$ (blue) and $i$ (red)  as a function of time,  compared with the deterministic solution, shown dashed, starting with one infected cervid and no prions at time $t=0$,   and (Right) $E(t)$ vs. $t$ for 50 different trials.  Parameter values are  $K=30$(100km$^{2}$)$^{-1}$, $A=2000$ km$^2$, $\mu_e = 0.2$yr$^{-1}$, $\rho_iW=0$, and other parameters as in Table \ref{table:parameters}. }
 \label{fig:stoch_1}
 \end{figure}
 
 \begin{figure}[ht]
\center
 \includegraphics[height=5cm]{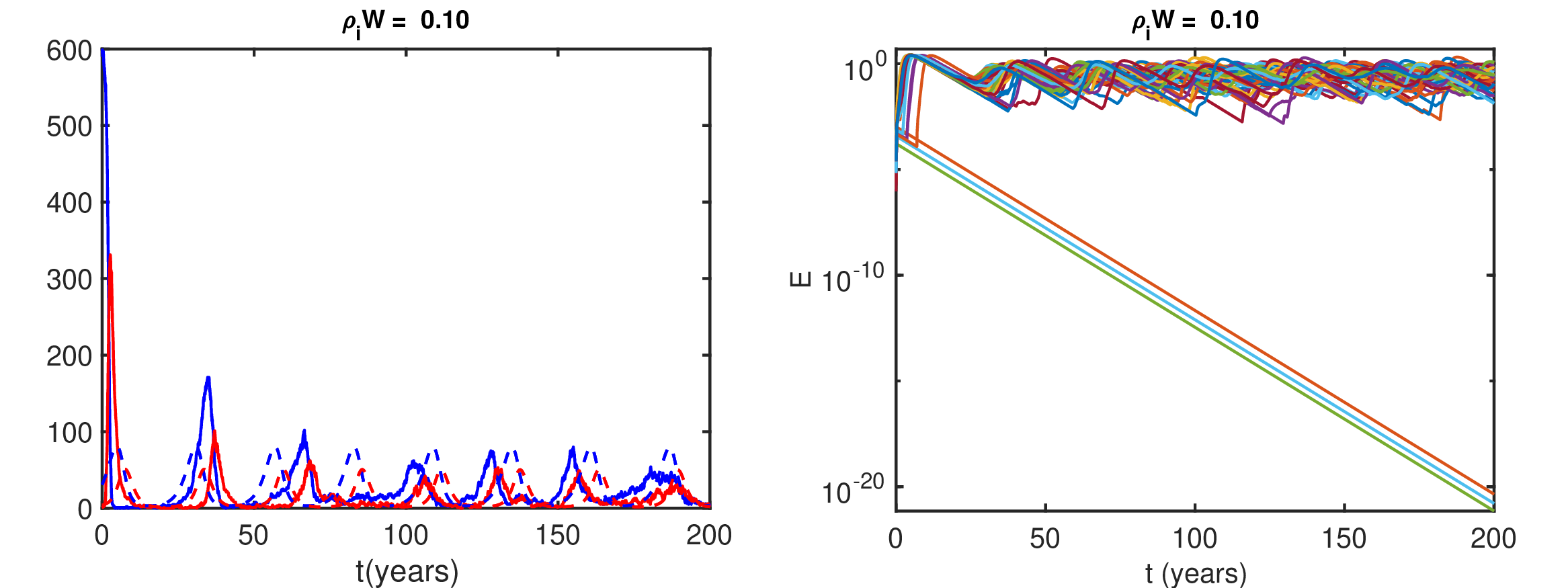}
  \caption{Example of stochastic simulation showing (Left) $s$ (blue) and $i$ (red)  as functions of time,  compared with the deterministic solution, shown dashed, starting with one infected cervid and no prions at time $t=0$,   and (Right) $E(t)$ vs. $t$ for 50 different trials.  Parameter values are  $K=30$(100km$^{2}$)$^{-1}$, $A=2000$ km$^2$, $\mu_e = 0.2$yr$^{-1}$, $\rho_iW=0.1$, and other parameters as in Table \ref{table:parameters}.}
 \label{fig:stoch_2}
 \end{figure}

 \begin{figure}[ht]
\center
 \includegraphics[height=5cm]{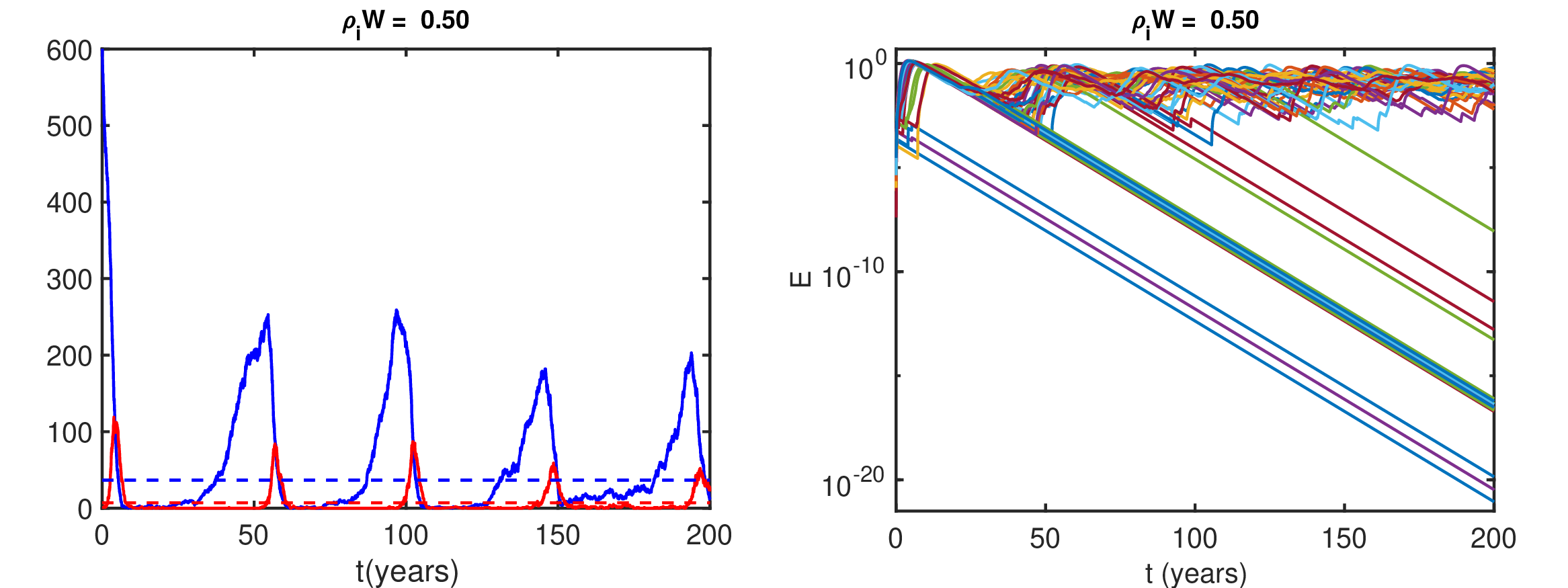}
  \caption{Example of stochastic simulation showing (Left) $s$ (blue) and $i$ (red)  as functions of time,  compared with the deterministic solution, shown dashed, starting with one infected cervid and no prions at time $t=0$,   and (Right) $E(t)$ vs. $t$ for 50 different trials.  Parameter values are  $K=30$(100km$^{2}$)$^{-1}$, $A=2000$ km$^2$, $\mu_e = 0.2$yr$^{-1}$, $\rho_iW=0.5$, and other parameters as in Table \ref{table:parameters}.}
 \label{fig:stoch_3}
 \end{figure}
 
  \begin{figure}[ht]
\center
 \includegraphics[height=5cm]{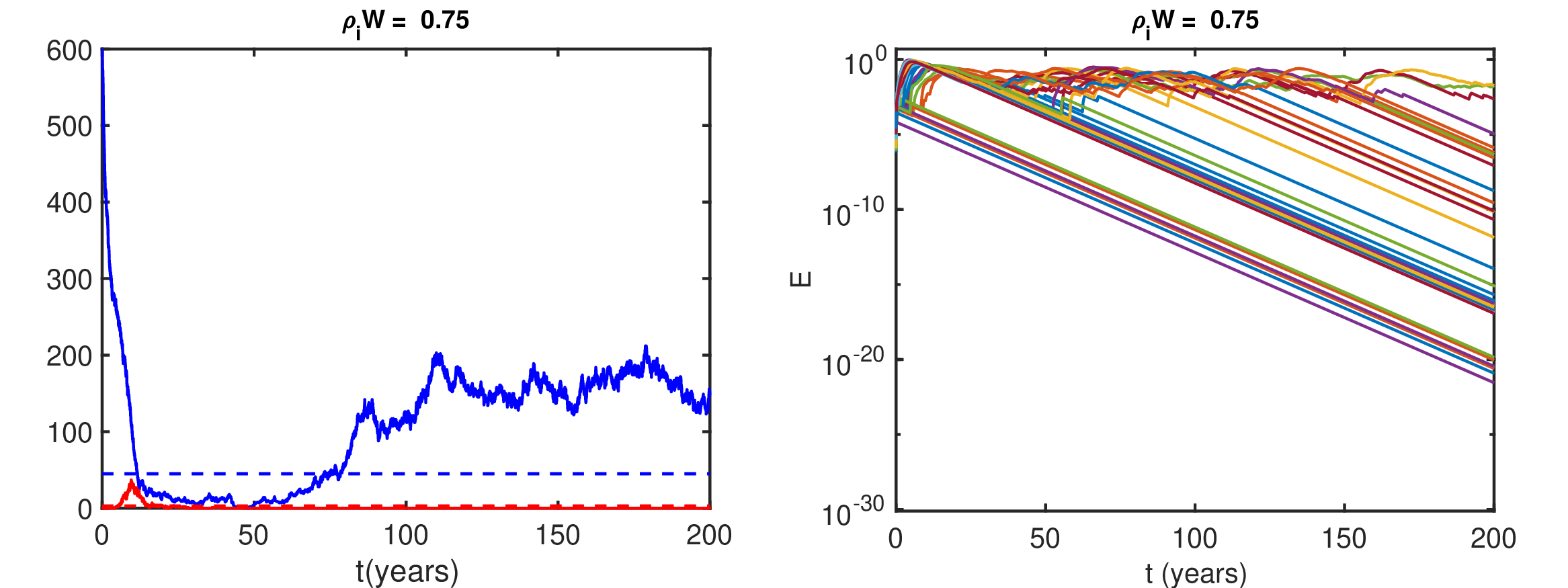}
  \caption{Example of stochastic simulation showing (Left) $s$ (blue) and $i$ (red)  as functions of time,  compared with the deterministic solution, shown dashed, starting with one infected cervid and no prions at time $t=0$,   and (Right) $E(t)$ vs. $t$ for 50 different trials.  Parameter values are  $K=30$(100km$^{2}$)$^{-1}$, $A=2000$ km$^2$, $\mu_e = 0.2$yr$^{-1}$, $\rho_iW=0.75$, and other parameters as in Table \ref{table:parameters}. }
 \label{fig:stoch_4}
 \end{figure}

The first stochastic simulation shown (Fig.~\ref{fig:stoch_1}) is with no predators ($\rho_iW=0$). For these parameter values, the deterministic model has oscillatory solutions. For this model with the stochastic simulation, the population $s$ always goes extinct in finite time \cite{childs_keener_keizer}, and so to prevent that from happening, $s$ is artificially never allowed to become zero. Our goal is to maintain stochastically oscillating populations to assess the possibility of controlling CWD outbreaks using wolves. However, it should be noted that the system cannot realize the extinction state $s=i=E=0$ when we make this modeling choice. Fig.~\ref{fig:stoch_1} (Right) shows the prion density trajectories for 50 independent trials, and it is seen here that most of the invasions with one infected cervid result in an oscillatory prion epidemic. The Figs.~\ref{fig:stoch_2}-\ref{fig:stoch_4} show a sample trajectory (Left)  and the prion density for 50 trajectories (Right) for the three values of $\rho_iW=0.1, 0.5$, and 0.75.

\begin{figure}[ht]
\center
 \includegraphics[height=5cm]{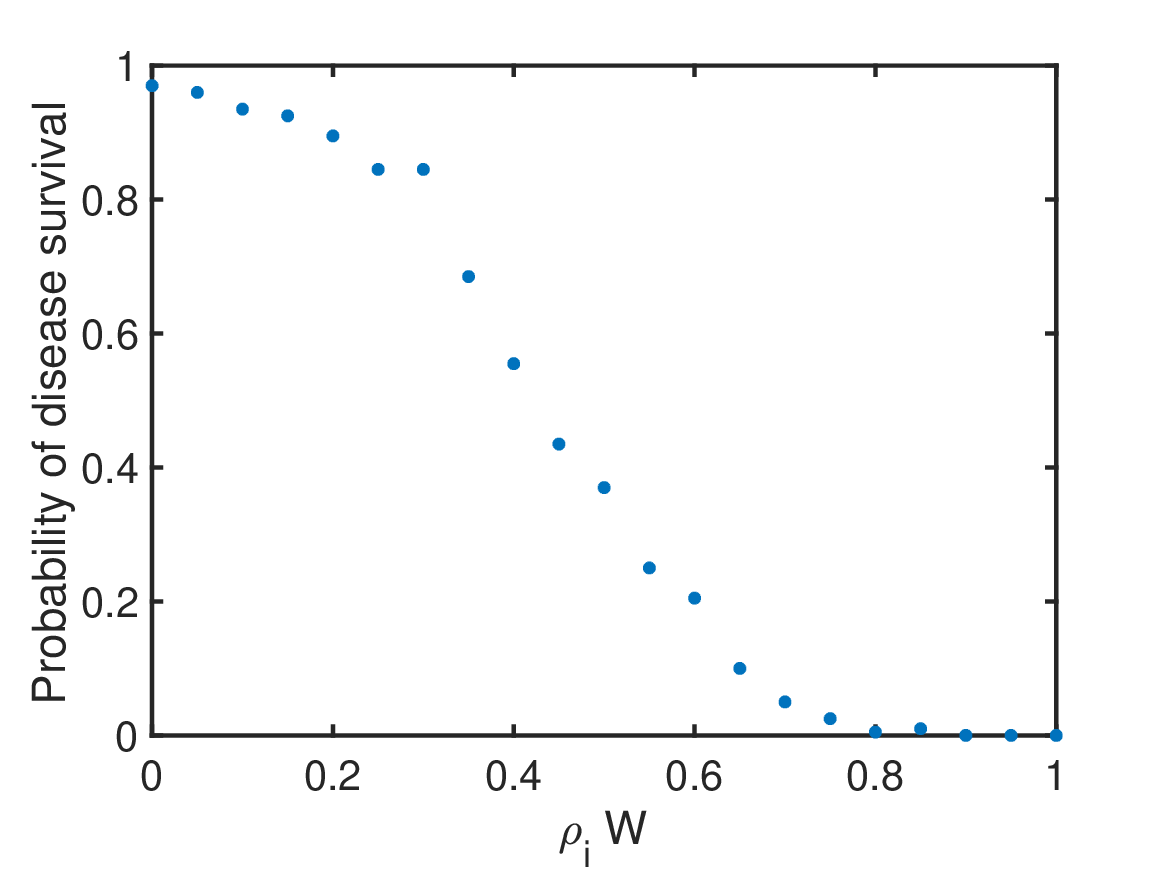}
  \caption{Probability of long-term survival of the disease following infection of a single individual, plotted as a function of $\rho_iW$. Parameters are as described in Table \ref{table:parameters}.}
 \label{fig:stoch_6}
 \end{figure}

There are several observations to make from these simulations.  First, the probability of an epidemic getting started from one infected individual is a decreasing function of the number of wolves present at the time of the appearance of the infected individual (see Figure \ref{fig:stoch_6}).  Second, it could be that the diseased animals are eradicated, but the disease reemerges because of the remnant prions in the environment.  However, third,   there is a possibility that the disease is permanently eradicated even after it initially spreads, and this probability is an increasing function of the number of wolves present.  For example, in Fig. \ref{fig:stoch_3}  roughly 80\% of the trajectories show disease elimination (with $\rho_iW=0.5$), while with $\rho_iW=0.75$ (Fig.~\ref{fig:stoch_4}) about  98\% of the trajectories show elimination of the disease.  We define disease eradication as $E < 0.001$ at the end of the simulation. This, of course, is somewhat arbitrary. To our knowledge, the smallest density of prions required for infection is currently unknown and may depend on the transmission pathway. See also related analytic calculations in Sections \ref{spread}-\ref{elimination}. More synthetic data for the probability of disease eradication is shown in Fig.~\ref{fig:stoch_6}, where the probability of long-term disease survival as a function $\rho_iW$ is shown, determined using 200 trajectories for each value of $\rho_iW$.

 \subsection{Disease Control}
 
 \label{control}
 
\begin{figure}[ht]
\center
 \includegraphics[height=5cm]{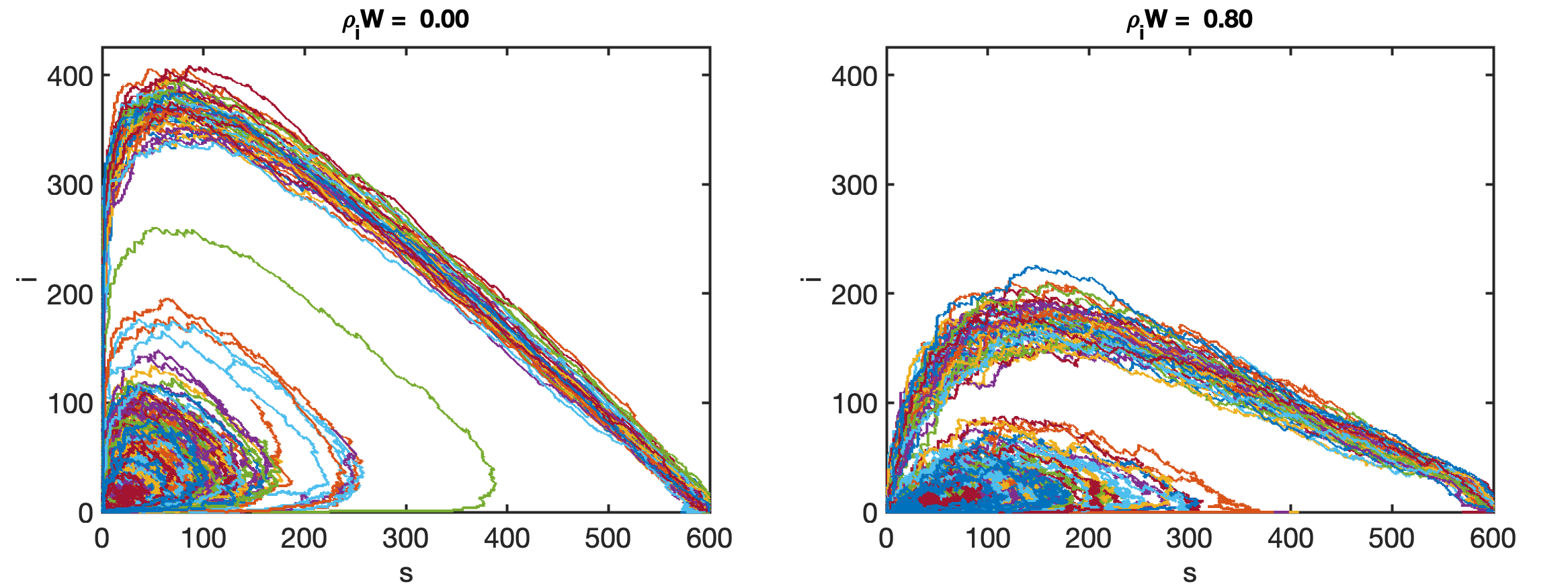}
  \caption{  $i$ vs $s$ ``phase portrait" of a CWD  infection with 50 different trials for which no intervention is made (Left) and (Right)  for which   $\rho_iW=0.8$ wolves are introduced when the infected numbers reach 30 (1/20$^{th}$ of the carrying capacity of 600).   
 Here  $K=30$(100km$^{2}$)$^{-1}$, $A=2000$ km$^2$, $\mu_e = 0.2$yr$^{-1}$, and other parameters as in Table \ref{table:parameters}.}
 \label{fig:stoch_intro_c}
  \end{figure}

\begin{figure}[ht]
\center
 \includegraphics[height=5cm]{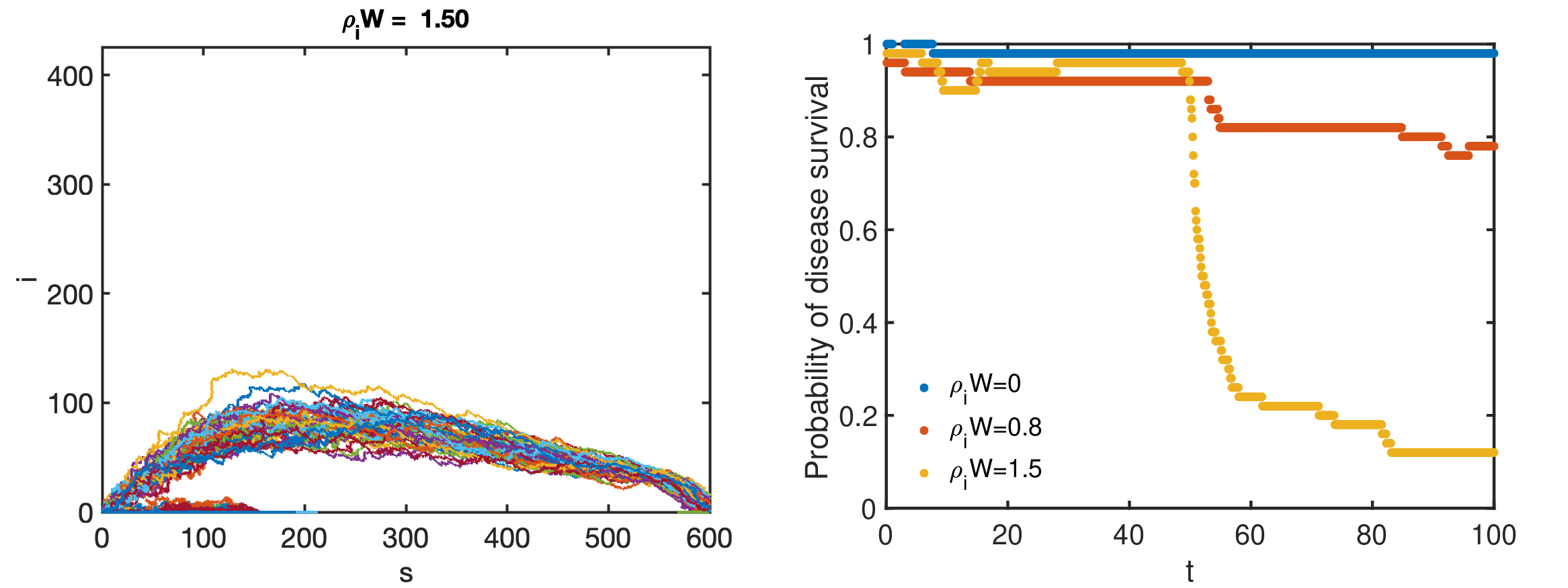}
  \caption{(Left) $i$ vs $s$ ``phase portrait" of a CWD  infection with 50 different trials for which intervention is made with  $\rho_iW=1.5$ wolves introduced when the infected numbers reach 30 (1/20$^{th}$ of the carrying capacity of 600).    (Right): Probability of disease survival as a function of time for the three different control scenarios, $\rho_iW=0.0$, $\rho_iW=0.8$, $\rho_iW=1.5$.
 Here  $K=30$(100km$^{2}$)$^{-1}$, $A=2000$ km$^2$, $\mu_e = 0.2$yr$^{-1}$, $\rho_sW=0.25 \rho_iW$, and other parameters as in Table \ref{table:parameters}.}
 \label{fig:stoch_intro_d}
  \end{figure}

For the previous simulations, it was assumed that there was a preexisting population of predators at the time the infected animal was introduced. Instead, it may be that wolves can be introduced only after the disease has been introduced and has begun to spread. In many regions\footnote{For example, see the 2024 Chronic Wasting Disease Surveillance Report from the Wyoming Game and Fish Department Report.}, the disease has already spread, and wildlife managers may wonder if introducing a population of predators could bring the disease under control.  Here, we attempt to model this situation by introducing wolves once the infected animals comprise 5\% of the population carrying capacity. The 5\% threshold was chosen, as we felt that it was high enough that wildlife managers could feasibly detect in their annual surveys, whereas lower thresholds may be challenging to recover in a statistically accurate manner from hunter-harvested deer. Certainly, other thresholds could be used. The result without the introduction of predators is shown in Fig.~\ref{fig:stoch_1}.  The result of this delayed introduction is summarized in Figs. \ref{fig:stoch_intro_c}-\ref{fig:stoch_intro_d}.  The phase portrait course of the disease for 50 trajectories with no intervention is shown in Fig.~\ref{fig:stoch_intro_c}(Left), and with the introduction of wolves with $\rho_iW=0.8$ in Fig.~\ref{fig:stoch_intro_c}(Right) and with $\rho_iW=1.5$ in Fig.~\ref{fig:stoch_intro_d}(Left).   In Fig.~\ref{fig:stoch_intro_d}(Right) is shown the probability of disease survival as a function of time for these three scenarios.  Apparent from these plots is that with the introduction of wolves, the severity of the disease is decreased in two ways, namely, the maximal level of infected animals is decreased, and the survival of uninfected animals is increased, giving improved opportunity for the herd to recover from the invasion.   

%
%

 \subsection{Probability of Spread}
 
 \label{spread}

We can get an analytical understanding of some of these results as follows:  Suppose there is one infected cervid with a death rate $\mu (=\mu_i+\rho_iW)$ introduced into a healthy population of size $s$.  What is the probability that prions shed by the one introduced infected individual and its interactions with healthy animals will lead to at least one other cervid becoming infected?  

The density of prions resulting from shedding of one infected cervid satisfies the differential equation $\frac{dE}{dt} = \frac{\epsilon A_0}{ A}-\mu_e E$ as long as the infected deer is alive $0<t<t_d$, and $\frac{dE}{dt} = -\mu_eE$ after its death.  
\beq
E(t) = \left\{\begin{array}{cc} \beta(1-\exp(-\mu_e t)),&0<t<t_d\\ \beta(1-\exp(-\mu_e t_d))\exp(-\mu_e(t-t_d),&t_d<t<\infty\end{array}\right.,
\label{eq:eqE}
\eeq
where $t_d$ is the time of death of the infected invader, and $\beta = \frac{\epsilon A_0}{\mu_e A}$.  
The probability $p_s$ of the disease spreading to at least one susceptible individual in a total population of $s$ by time $t$ is given by (ignoring all other births or deaths) 
\beq
\frac{dp_s}{dt} = \Big(\gamma_e s E(t)+\gamma_i{A_0\over A}s(i==1)\Big)(1-p_s),
\eeq
from which it follows that the probability of spread by time $t$ is given by 
\beq
p_s(t) = 1-\exp\Big(-\gamma_e s \int_0^t E(\sigma) d\sigma -\gamma_is{A_0\over A}(t_d-H(t_d-t)(t-t_d))\Big),
\eeq
where $H$ is the Heaviside function, and the probability of eventual spread is
\beq
p_s(\infty) = 1-\exp\Big(-\gamma_e s \int_0^\infty E(\sigma) d\sigma\ -\gamma_is{A_0\over A}t_d\Big).
\eeq
A direct calculation using (\ref{eq:eqE}) yields
\beq
p_s(\infty) = 1-\exp\Big(-\gamma_e s \beta t_d-\gamma_is{A_0\over A}t_d\Big).
\eeq
Now, the probability that the infected animal dies at time $t_d$ is given by the probability density function $p_d(t_d) =\mu\exp(-\mu t_d)$, so the probability of spread, taking into account that the time of death $t_d$ is random, is
\beq
p_s =\mu \int_0^\infty \Big(1-\exp (-\gamma_e s \beta t_d-\gamma_i{A_0\over A}t_d)\Big) \exp(-\mu t_d) dt_d   = \frac{1}{\frac{\mu}{\gamma_e s \beta+\gamma_is{A_0\over A}}+1}.\label{eq:prob_of_spread}
\eeq
In terms of original parameters,
\beq
\frac{\mu}{\gamma_e s \beta+\gamma_is{A_0\over A}} = \frac{A\mu_e(\mu_i+\rho_iW)}{sA_0(\gamma_e  \epsilon  +\gamma_i\mu_e)}.
\eeq

Notice that $p_s$ in (\ref{eq:prob_of_spread}) is a decreasing function of $\mu$ and an increasing function of $s$.  
Thus,  predation of infected cervids is a beneficial thing, since anything that can increase $\mu$, the death rate of infected cervids, will decrease the probability of spread of the infection.   In other words, the presence of wolves in a disease-free population helps to prevent the invasion of the disease by an immigrant infected animal.  Further, observe that overpopulation of susceptible cervids increases the probability of the spread of the disease, while predation of the susceptible population, which decreases $s$ through $\rho_sW$, helps prevent the spread of the disease.  Thus, the presence of predators is beneficial for at least these two reasons.

\drop{It is interesting  to note that this formula is {\it exactly} the same in form as for the case of an SIR disease with infection mediated by direct contact between the infecteds and susceptibles.  Further, the parameter $R_0 = \frac {\gamma_e s \beta}{\mu}$ is the classical parameter for the threshold of spread. 
}

 \subsection{Probability of Elimination}
 
 \label{elimination}
 
Because prions survive in the environment, even if all infected animals are eliminated by some control mechanism, there is the possibility that the disease will reemerge some time later.  To calculate this probability,   let $p_{s,0}(t)$ be the probability that there are $s$ susceptible deer and no infecteds at time $t$, but there is some contamination of the environment with prions, $E$.  It follows that
\beq
\frac{dp_{s,0}}{dt} = -\gamma_esE(t)p_{s,0}.
\eeq
(Here, again, we are ignoring births and deaths of healthy animals). $E(t)$, of course, is the density of prions in the environment, which, as long as there are no infecteds, satisfies $\frac{dE}{dt} = -\mu_eE(t)$.
Consequently, 
\beq
E(t)=E_0\exp(-\mu_et),
\eeq
where $E_0$ is the initial density of prions at time $t=0$, so that 
\beq
\frac{dp_{s,0}}{dt} = -\gamma_esE_0\exp(-\mu_et)p_{s,0}.
\eeq
It follows that
\beq
\ln(p_{s,0}(t)) = {\gamma_e\over \mu_e}sE_0(\exp(-\mu_et)-1),
\eeq
and
\beq
\lim_{t\rightarrow\infty}p_{s,0} = \exp(-{\gamma_e\over \mu_e}sE_0)
\eeq
is the probability that there will not be another outbreak.  Obviously, the probability that there will not be another outbreak goes to zero as ${\gamma_e\over \mu_e}sE_0$ becomes large.  This once again shows the risk associated with overpopulation of susceptible deer (large $s$), and the benefit of having wolves or other harvesting measures to help control the population size of healthy deer, even when there are no infected animals present. This is borne out by a recent statistical study that showed that deer populations that experienced high harvest pressure had lower prevalence of CWD \cite{moss2025effectiveness}. 

\section{Control in the full dynamical model}


\label{siew}

In our analysis so far, we took $W$ to be constant, to keep the analysis more analytically tractable (see Section \ref{bif}). Here, we explore the control implications of CWD outbreaks when $W$ is dynamic. Consider the following augmented form of Eqs. (\ref{eq:m7a})-(\ref{eq:m9})

\bea
\frac{dS}{dt} &=& rS\left(1-\frac{S+ I}{K}\right)-\gamma_e SE -\gamma_i SI-\rho_sSW,\label{eq:s}\\
\frac{dI}{dt} &=& \gamma_eSE+\gamma_i SI-\mu_i I-\rho_iIW = \gamma_eSE+\gamma_i SI-\mu_i I-\omega \rho_sIW, \label{eq:i}\\
\frac{dE}{dt} &=& \epsilon I -\mu_e  E \label{eq:e},\\
\frac{dW}{dt} &=& b_w W\left(1 - \dfrac{W}{K_w'}\right) -d_w W+ \alpha(\rho_sSW+ \rho_iIW)=r_{w}W\left(1 - \dfrac{W}{K_w}\right) + \alpha(\rho_sSW+ \omega \rho_sIW). \label{eq:w}
\ena

Here, we assume that $W$ grows at some rate $r_w$ and the population is governed by a carrying capacity, $K_w$. As in Eq. (\ref{eq:m7a}), we take $r_w = b_{w} -d_{w} $, where  $b_{w}$ is the  $d_w$ are zero population birthrate for wolves and the natural mortality rate for wolves, respectively. We take $b_w=0.5$ and $d_w=0.2$, as we thought it reasonable that wolves should have a lower birthrate than deer but a higher natural mortality rate than deer. As before in Eq. (\ref{eq:m7a}) , $K_w = K_w' (1 - \dfrac{d_w}{b_w})$.  We selected the carrying capacity to be $K_w=3$ (100 km$^{2})^{-1}$. Clearly, it must be true that $K_w < K$, and this value seemed reasonable with estimates of wolf densities \cite{mech2019wolves}, though the reported densities have a large spread, and other choices for this parameter could have been made. We also assume predation increases the population size, through some conversion efficiency, $\alpha$, but very weakly. We assume that $\alpha <<1$, as most predation events should not significantly contribute to the growth of the wolf population. In Section \ref{bif}, we assumed that $\rho_s < \rho_i$, and set $\frac{\rho_s}{\rho_i} = 0.5$, as predators may selectively target weaker prey. Here, we set $\rho_i = \omega \rho_s$, where $\omega \geq 1$ (Eqs. (\ref{eq:s})-(\ref{eq:w})). Biologically, $\omega$ represents the degree of the predation bias towards infected deer over susceptible deer. We will treat $\omega$ as our bifurcation parameter. At steady state, Eqs. (\ref{eq:s})-(\ref{eq:w}) have five physical equilibria: 

{\normalsize
\begin{align}
	& S=I=E=W=0, \\
	& S=K, I=E=W=0, \\
	& S= -\frac{K r_w \left(K_w  \rho_s - r \right)}{K K_w \alpha  \rho_s^{2}+r r_w}, I=E=0, W=\frac{K_w r \left(K \alpha  \rho_s +r_w \right)}{K K_w \alpha  \rho_s^{2}+r r_w}, \label{abol} \\
	& S=\frac{\mu_e \mu_i}{\epsilon  \gamma_e+\gamma_i \mu_e}, I= \frac{\mu_e r \left(K \epsilon  \gamma_e+K \gamma_i \mu_e-\mu_e \mu_i\right)}{\left(\epsilon  \gamma_e+\gamma_i \mu_e\right) \left(K \epsilon  \gamma_e+K \gamma_i \mu_e+\mu_e r\right)}, E=\dfrac{\epsilon}{\mu_e}I, W=0,\\
	& S=ss_s, I= ss_i, E=\dfrac{\epsilon}{\mu_e} ss_i, W=ss_w, \label{endem}
\end{align}
}

where 

$$ss_s =\frac{ \mu_e  \left(\left(\left( K_w  \omega  \rho_s + \mu_i \right) \left(K  \gamma_i +r\right)  r_w +K \alpha  \omega   K_w  \rho_s^{2} \left(r \omega + \mu_i \right)\right)  \mu_e +K \epsilon   \gamma_e   r_w  \left( K_w  \omega  \rho_s + \mu_i \right)\right)}{\left( \gamma_i  \left(K  \gamma_i +r\right)  r_w +\alpha  \omega  r  K_w  \rho_s^{2} \left(\omega -1\right)\right)  \mu_e ^{2}+2 \epsilon   r_w   \gamma_e  \left(K  \gamma_i +\frac{r}{2}\right)  \mu_e +K \epsilon^{2}  \gamma_e ^{2}  r_w }, $$

$$ ss_i = \frac{ \mu_e  \left(\left(\left(- \gamma_i  \left(- K_w  \rho_s +r\right) K+r \left( K_w  \omega  \rho_s + \mu_i \right)\right)  r_w +K \alpha   K_w  \rho_s^{2} \left(r \omega + \mu_i \right)\right)  \mu_e -K \epsilon   \gamma_e   r_w  \left(- K_w  \rho_s +r\right)\right)}{\left( \gamma_i  \left(K  \gamma_i +r\right)  r_w +\alpha  \omega  r  K_w  \rho_s^{2} \left(\omega -1\right)\right)  \mu_e ^{2}+2 \epsilon   r_w   \gamma_e  \left(K  \gamma_i +\frac{r}{2}\right)  \mu_e +K \epsilon^{2}  \gamma_e ^{2}  r_w },$$

$$ss_w =\frac{ K_w  \left(z_1(\gamma_i, K, r, r_w, \alpha, \rho_s, \omega, \mu_i)+z_2(\gamma_e, K, \gamma_i, r, r_w, \alpha, \rho_s, \mu_i, \mu_e,\epsilon) \right)}{\left( \gamma_i  \left(K  \gamma_i +r\right)  r_w +\alpha  \omega  r  K_w  \rho_s^{2} \left(\omega -1\right)\right)  \mu_e ^{2}+2 \epsilon   r_w   \gamma_e  \left(K  \gamma_i +\frac{r}{2}\right)  \mu_e +K \epsilon^{2}  \gamma_e ^{2}  r_w },$$

where 

$$z_1(\gamma_i, K, r, r_w, \alpha, \rho_s, \omega, \mu_i)= \left( \gamma_i  \left(K  \gamma_i +r\right)  r_w +\alpha  \rho_s  \left( \gamma_i  \left(r \omega + \mu_i \right) K-r  \mu_i  \left(\omega -1\right)\right)\right)  \mu_e ^{2}, $$

and 

$$z_2(\gamma_e, K, \gamma_i, r, r_w, \alpha, \rho_s, \mu_i, \mu_e,\epsilon) = \epsilon   \gamma_e  \left(\left(2 K  \gamma_i +r\right)  r_w +K \alpha  \rho_s  \left(r \omega + \mu_i \right)\right)  \mu_e +K \epsilon^{2}  \gamma_e ^{2}  r_w.$$



  \begin{figure}[H]
\center
 \includegraphics[height=15cm]{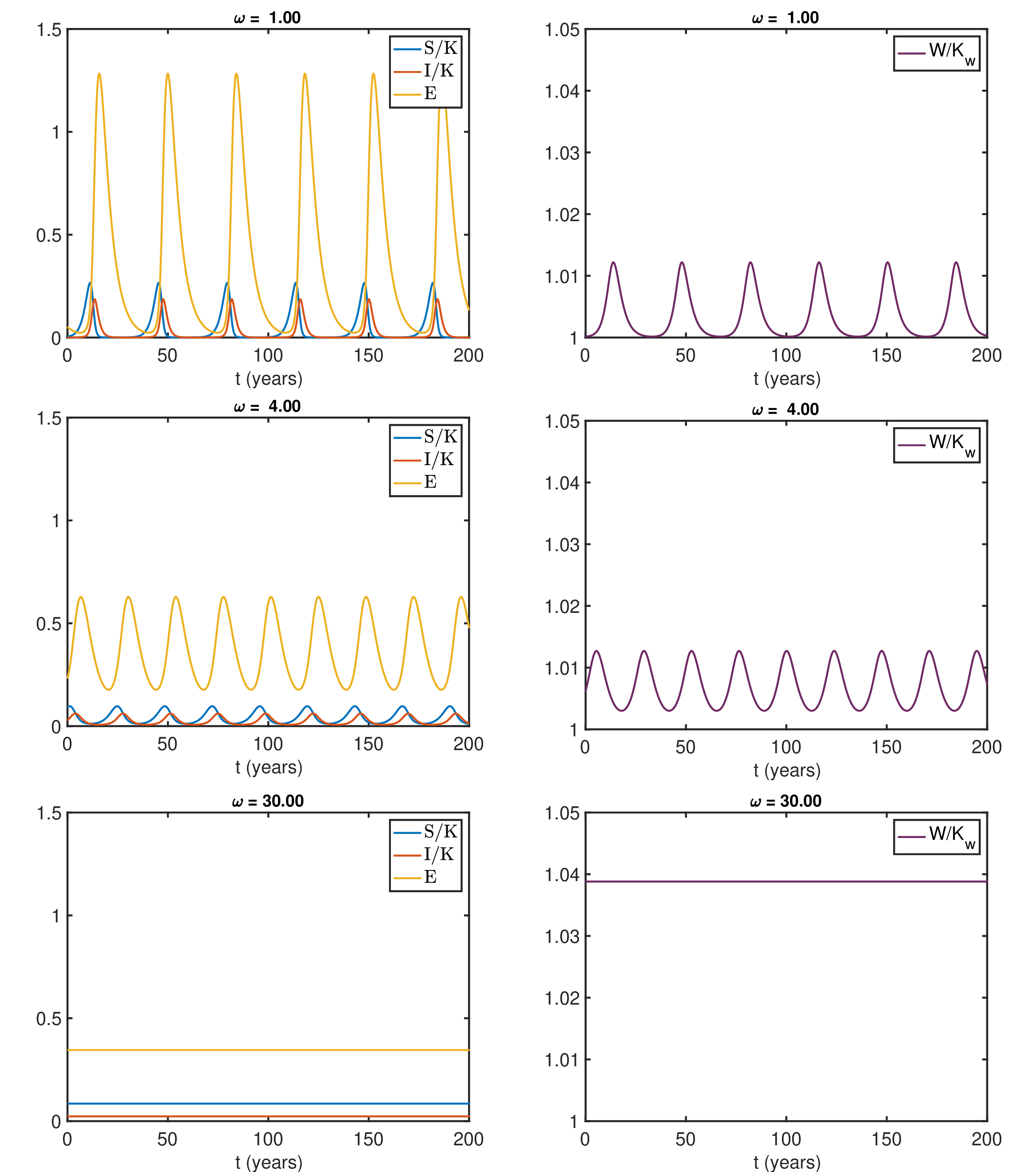}
  \caption{Wolves may be able to be used to control CWD outbreaks, assuming a fully dynamic model (Eqs. (\ref{eq:s})-(\ref{eq:w})). Here we simulate Eqs. (\ref{eq:s})-(\ref{eq:w}) deterministically for variable $\omega$. The parameter $\omega = \dfrac{\rho_i}{\rho_s}$ describes the ratio of the predation rate of infected deer to the predation rate of susceptible deer. Top: oscillatory CWD outbreaks are possible when $\omega =1$, which, biologically, occurs when the predation rates on $S$ and $I$ are equal, meaning that the predator exhibits no preference for infected deer over susceptible deer. Middle: Increasing the selective predation of the infected deer ($\omega=4$) dampens oscillations in the system. Bottom: Further increasing $\omega$ drives the deterministic system to the stable endemic state, similar to the constant case (see Figure \ref{fig:3de}). Here, we take $\rho_s = 0.01$, $\rho_i = \omega \rho_s$, $\alpha=0.05$, $r_w = 0.3$, $K_w=3$, and all other parameters as listed in Table \ref{table:parameters}.}
 \label{fig:dyn_w_det}
  \end{figure}

Due to the nonlinear structure of Eqs. (\ref{eq:s})-(\ref{eq:w}), computing analytic expressions that describe the location of the Hopf bifurcation in parameter space becomes intractable. The analytic expression becomes so large and unwieldy that computer algebra systems struggle to compute the curve. However, we can still investigate the system numerically to determine if predators can be used to control a CWD outbreak when the system is governed by Eqs. (\ref{eq:s})-(\ref{eq:w}). We see that Eqs. (\ref{eq:s})-(\ref{eq:w}) emit an oscillatory solution, which biological represents oscillatory CWD outbreaks, for the parameter values listed in Table \ref{table:parameters}, $\rho_s = 0.01$, $r_{w}=0.3$, $K_w=3$, and sufficiently low $\alpha$ and $\omega$ ($\alpha = 0.05$, and $\omega=1$, see Figure \ref{fig:dyn_w_det}, top). Increasing $\omega$ dampens the oscillation (see Figure \ref{fig:dyn_w_det}, middle), and further increasing $\omega$ drives the system into the endemic state (Eq. \ref{endem}) where oscillatory outbreaks are eliminated in the deterministic model. Computationally, the transition to the endemic state occurs around $\omega \approx 5$ for the parameter values we used in our study. Moving the system from oscillatory CWD outbreaks to the endemic state comes at the cost of reducing the overall deer population well below the carrying capacity. It may be possible for wolves or other predators to control or at least temper CWD outbreaks, as we see here, but $\omega$ needs to be sufficiently high for this strategy to work. Biologically, this means that predators need to selectively prey upon infected deer at a sufficiently high rate as compared with the predation rate of susceptible deer. As before, we also study a stochastic version of the system, as the populations are expected to be relatively small. 

We stochastically simulate the system for variable $\omega$, using the modified Gillespie algorithm (see Section \ref{stoch}). As wolves are no longer constant, we need to modify existing reactions and include several new reactions. In particular, we have

\bea
R_1: &&s\rightarrow s+1 \quad {\rm (Birth~of~deer)}\\
R_2: &&s\rightarrow s-1 \quad {\rm (Death~of~deer)}\\
R_3:&&s, i\rightarrow s-1,i+1 \quad {\rm (Infection~of~deer~ by ~E)}\\
R_4:&&s,  i\rightarrow s-1,i+1 \quad {\rm (Infection~of~deer~ by~}i)\\
R_5:&&i\rightarrow i-1\quad {\rm (Death~of~infected~deer)}\\
R_6:&&s \rightarrow  s-1 \quad {\rm (Predation ~of~uninfected~deer~no~birth~of~wolf)}\\
R_7:&&i \rightarrow i-1  \quad {\rm (Predation ~of~infected~deer~no~birth~of~wolf)} \\
R_8:&& s,w \rightarrow s-1,w+1 \quad {\rm (Predation ~of~uninfected~deer~and~birth~of~wolf)} \\
R_9:&&  i,w \rightarrow i-1,w+1 \quad {\rm (Predation ~of~infected~deer~and~birth~of~wolf)}\\
R_{10}:&& w \rightarrow w+1 \quad {\rm (Birth~of~wolf)} \\
R_{11}: && w \rightarrow w-1 \quad {\rm (Death~of~wolf)} 
\ena

where $s$ and $i$ represent the number of susceptibles and infected, respectively, at any given time, as before, and similarly, $w$ represents the number of wolves at any given time. These reactions occur at the rates:

\bea
R_1:&& r_1 = b s(1-{A_0\over K' A}(s+i)), \label{bs_rate}\\
R_2: &&r_2 =  d s,\\
R_3:&& r_3 = \gamma_e E s,\\
R_4:&& r_4 = \gamma_i {A_0\over A}si,\\
R_5:&& r_5 = \mu_i i,\\
R_6:&&r_6= (1-\alpha) \rho_s  {A_0\over A} sw,\\
R_7:&&r_7= (1-\alpha) \rho_i {A_0\over A} iw, \\
R_8:&&r_8=\alpha \rho_s  {A_0\over A} sw\\
R_9:&&r_9=\alpha \rho_i {A_0\over A} iw\\
R_{10}:&&r_{10}=b_{w} w(1 - {A_0\over K_w' A}w) \label{ws_rate}\\
R_{11}: &&r_{11}=d_{w} w
 \ena

\begin{figure}[H]
\center
 \includegraphics[height=15cm]{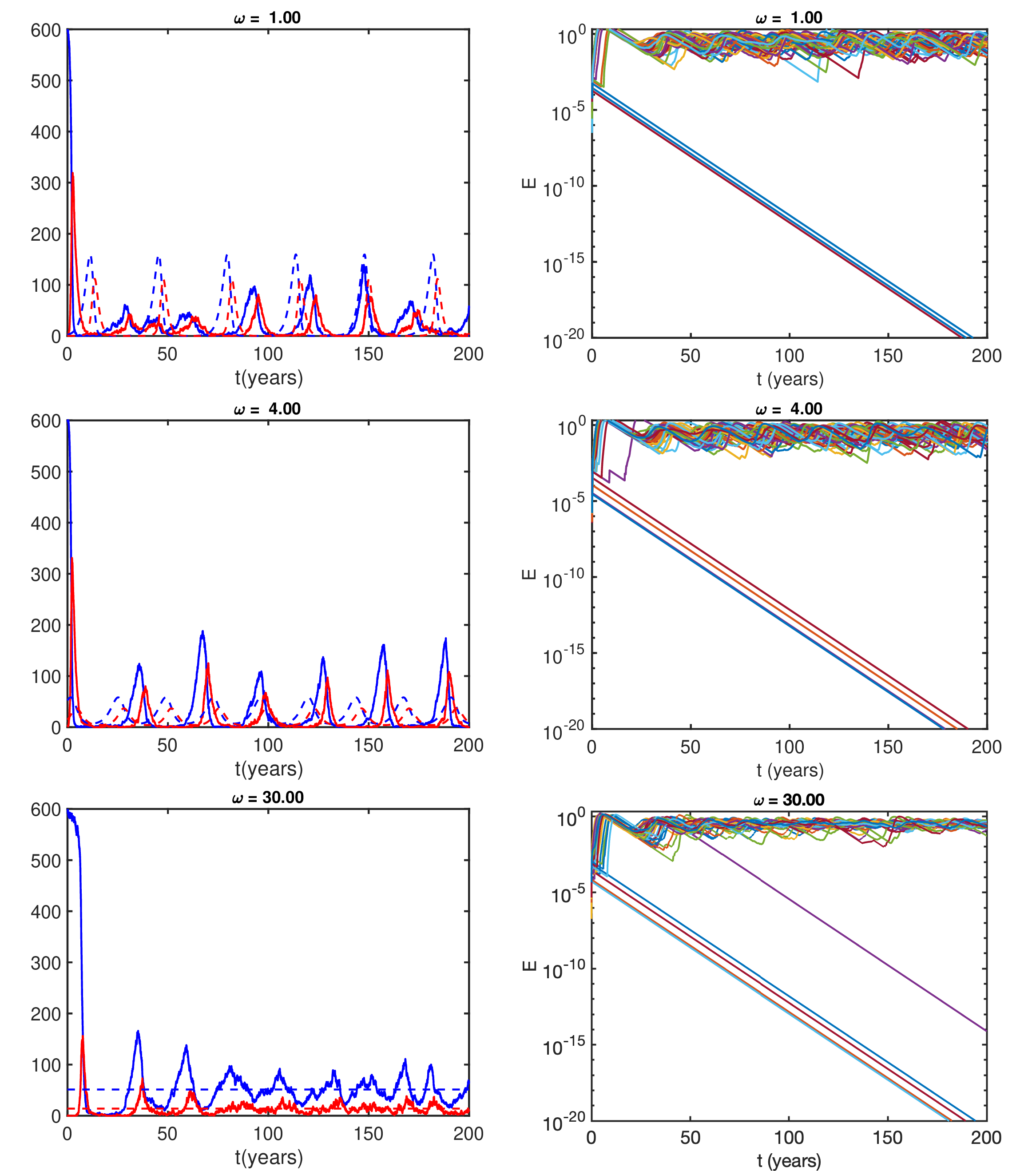}
  \caption{Wolves may be able to be used to control CWD outbreaks, assuming that $\omega$ is sufficiently high. Top: oscillatory CWD outbreaks when $\omega=1$. Middle: Dampened oscillations when $\omega=4$. Bottom: Many trajectories move to the endemic state (Eq. \ref{endem}) for $\omega=30$. Left: Example of stochastic simulation of Eqs. (\ref{eq:s})-(\ref{eq:w}) showing $s$ (blue) and $i$ (red)  as functions of time, compared with the deterministic solution, shown dashed, starting with one infected cervid and no prions at time $t=0$. Right: $E(t)$ vs. $t$ for 50 different trials. Here, we take $\rho_s = 0.01$, $\rho_i = \omega \rho_s$, $\alpha=0.05$, $r_w = 0.3$, $K_w=3$, and all other parameters as listed in Table \ref{table:parameters}. Predator dynamics not shown.}
 \label{fig:dyn_w_stoch}
  \end{figure}

\begin{figure}[H]
\center
 \includegraphics[height=5cm]{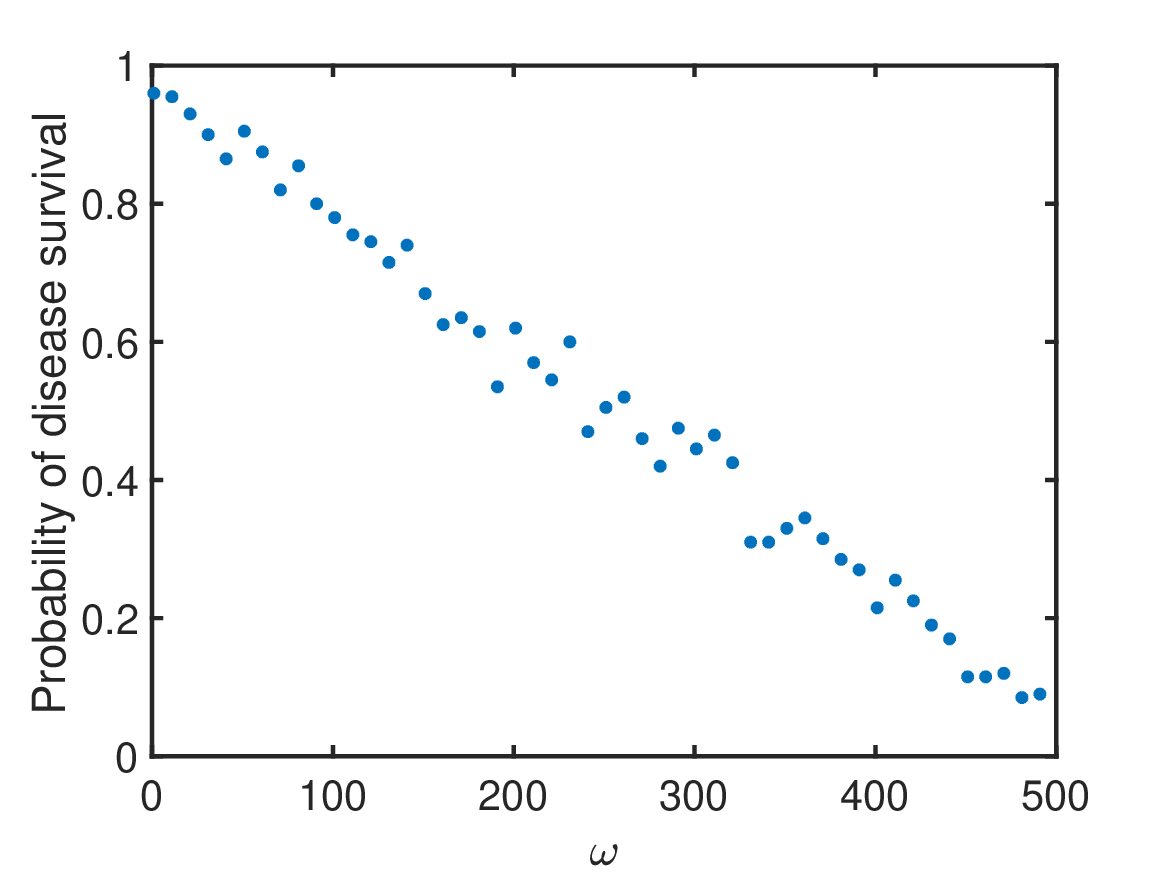}
  \caption{Probability of long-term survival of the disease following infection of a single individual in a stochastic simulation of Eqs. (\ref{eq:s})-(\ref{eq:w}),
plotted as a function of $\omega$. 200 trajectories were used for each value of $\omega$. Here, we set $\rho_s = 0.01$, $\rho_i = \omega \rho_s$, $\alpha=0.05$, $r_w = 0.3$, $K_w=3$, and all other parameters as listed in Table \ref{table:parameters}.}
 \label{fig:elim_prob_omega}
  \end{figure}

As before (see Sections \ref{bif}-\ref{stoch}), we see that it is possible to use wolves to force the system into the endemic disease state, but the effectiveness of the strategy crucially depends on the parameter $\omega$, which describes the selectively of predation on the infected deer population (see Figure 
\ref{fig:dyn_w_stoch}). We also see that when the predator is assumed to be dynamic, it may be less likely that wolves eradicate the disease completely, and instead we may expect wolves to push the system into the endemic state (Eq. \ref{endem}) rather than the disease-free state (Eq. \ref{abol}) (see Figure \ref{fig:elim_prob_omega}). To ensure that the disease is eradicated when the predator is dynamic requires $\omega >> 1$, which may not be biologically reasonable (see Figure \ref{fig:elim_prob_omega}). Further biological research is needed to determine estimates of $\omega$. 

\section{Discussion}


\label{Discussion}
 
 Here, we performed a bifurcation analysis of a simple deterministic mathematical model (Eq. (\ref{eq:m7})-(\ref{eq:m9})) that describes the direct and indirect transmission of CWD in a population of deer subject to predation by wolves, which we originally assume are constant (see Figures \ref{fig:1de}-\ref{fig:3de}, Section \ref{bif}). We then stochastically simulated the system using the Gillespie Algorithm and applied Poisson-thinning to the prion density, as it varies continuously in time (see Figures \ref{fig:stoch_1}-\ref{fig:stoch_6}). This model investigation demonstrates that introducing predators, such as wolves, may be an effective way to control diseases such as CWD, for which there are no other known effective control strategies. We also demonstrated that this control strategy works even if CWD is already prevalent in the population, as is the case for many cervid populations in the American West and Upper Midwest (see Figures \ref{fig:stoch_intro_c}-\ref{fig:stoch_intro_d}, Section \ref{control}). We also compute the probability of the disease spreading to at least one susceptible individual in a total population of $s$ by time $t$ (Section \ref{spread}) and the probability that no future outbreak will occur given a completely susceptible population and $E$ in the environment (Section \ref{elimination}). These calculations make clear the need to control not only the size of the infected population, but also the size of the susceptible population. 
 
 Finally, we study a more complex version of the system (Eqs. (\ref{eq:s})-(\ref{eq:w})) in which the predator is assumed to be dynamic, rather than constant, as in (Eqs. (\ref{eq:m7})-(\ref{eq:m9})). Our analysis of the fully dynamic model (Eqs. (\ref{eq:s})-(\ref{eq:w})) shows that wolves still may be effective for controlling CWD, but the strategy depends on strong selective predation of infected deer over susceptible deer, which we describe using the parameter $\omega$ (see Figures \ref{fig:dyn_w_det}, \ref{fig:dyn_w_stoch}, and Section \ref{siew}). For the parameter regime we used, we found that full eradication of the disease required $\omega>>1$, which may not be biologically reasonable, and so one might expect that introducing wolves may ultimately push the system towards the endemic state (Eq. \ref{endem}) rather than the disease-free state (Eq. \ref{abol}). Further biological research is needed to determine estimates of $\omega$, and specifically if $\omega > 1$, though some evidence supports the idea that some predators may selectively target infected deer at least compared with human hunters \cite{krumm2010mountain}. A recent study using a more complex, age-structured deterministic model also found that predators may help to limit CWD outbreaks, but the result also depended on predation selectivity \cite{brandell2022examination}. Taken together, our analysis highlights wolves, or other predators, as a possible effective control strategy for CWD and makes clear the role for human-hunting or other harvesting measures to limit the susceptible population size. 
 
Our analysis has several limitations relating to the parameter values, model complexity, and modeling assumptions. We found it somewhat difficult to find published parameter values to parameterize our mathematical models with. Ultimately, we used parameter estimates from \cite{miller_2006} and previous modeling efforts \cite{SharpPastor2011}, and otherwise used intuition to make reasonable guesses about the magnitude of the parameters. We noticed that some of the models used to generate parameter estimates may lack structural identifiability \cite{miller_2006}. Consider the ``Indirect + Direct'' model examined in \cite{miller_2006}, which has a similar form to Eq. (\ref{eq:m7})-(\ref{eq:m9}). The ``Indirect + Direct'' model has the form
 
\begin{align*} 
\text{Number of susceptible deer, Unobserved}~\dfrac{dS}{dt} &= a - S(\beta I + \gamma E + m), \\
\text{Number of infected deer, Unobserved}~\dfrac{dI}{dt} &= S(\beta I + \gamma E) - I(m + \mu), \\
\text{Mass of infectious material in the environment, Unobserved}~\dfrac{dE}{dt} &= \epsilon I - \tau  E, \\
\text{Number of dead infected deer, Observed}~\dfrac{dC}{dt} &= \mu I.
\end{align*} 
 
 The authors attempted to fit this model using observations of $C$ from two data sets of CWD outbreaks. However, notice that the system has a scaling symmetry \footnote{We used \texttt{STRIKE-GOLDD} to find the scaling symmetry \cite{massonis2020finding}. The MATLAB implementation we used can be found here \url{https://github.com/afvillaverde/strike-goldd}. The ``Indirect'' model in \cite{miller_2006} may also have the same scaling symmetry. Other models were not examined.} , which causes a structural identifiability issue (see $\textcolor{blue}{\psi}$ below). The structure of the model prohibits the discovery of a unique parameter estimate because the dynamics of the system are the same for any choice of $ \textcolor{blue}{\psi}$. 
 
 \begin{align*} 
\text{Number of susceptible deer, Unobserved}~\dfrac{dS}{dt} &= a - S(\beta I + \dfrac{\gamma}{\textcolor{blue}{\psi}} \textcolor{blue}{\psi} E + m), \\
\text{Number of infected deer, Unobserved}~\dfrac{dI}{dt} &= S(\beta I + \dfrac{\gamma}{\textcolor{blue}{\psi}} \textcolor{blue}{\psi} E) - I(m + \mu),  \\
\text{Mass of infectious material in the environment, Unobserved}~\textcolor{blue}{\psi} \dfrac{dE}{dt} &= \textcolor{blue}{\psi} \epsilon I - \tau \textcolor{blue}{\psi} E,\\
\text{Number of dead infected deer, Observed}~\dfrac{dC}{dt} &= \mu I.
\end{align*} 
 
 A structural identifiability analysis \footnote{We used \texttt{StructuralIdentifiability.jl} to perform the identifiability analysis of the system \cite{structidjl}. The implementation we used can be found here \url{https://docs.sciml.ai/StructuralIdentifiability/stable/}. For a structural identifiability tutorial in an epidemiological context, see \cite{massonis2021structural}. } of the model structure reveals that the parameters $\gamma$ and $\epsilon$ are structurally non-identifiable when only $C$ is observed, though the other kinetic parameters are globally identifiable. A recent movement ecology analysis showed that susceptible and infected deer can be detected from GPS tracks \cite{barrile2024chronic}, possibly providing a cheap method for temporal measurement of $S$ and $I$. Given this finding and temporal estimates of $S$ and $I$ from state CWD surveillance programs, it is tempting to think that if one could measure $S$, $I$, and $C$, instead of just $C$, the structural identifiability issue may be resolved. This is not true. The parameters $\gamma$ and $\epsilon$ are still globally non-identifiable even if $S$, $I$, and $C$ are observed, but $E$ is unobserved. However, if one is able to observe $C$ and $E$, but not $S$ and $I$, all kinetic parameters are globally identifiable, but measuring $E$ is a key challenge for the system. The prions responsible for CWD are known to exist in soil, and can be measured from soil samples \cite{kuznetsova2024detection}. In practice, estimating $E$ through soil samples would likely be expensive, infeasible on a large scale, and may miss some of the sources of the prions in the environment. The prions responsible for CWD can be found in multiple environmental reservoirs beyond soil.  It was recently shown that ticks can harbor levels of prions that could cause transmission of CWD \cite{inzalaco2023ticks}. Prions were also recently shown to exist in plants at relevant levels for transmission of CWD \cite{carlson2023plants}. For these reasons, getting estimates of the prion density will be difficult at best or otherwise infeasible. 
 
 An alternative approach to ensure that unique parameters can be found is to reparameterize the model to remove the structural identifiability issue ahead of parameter estimation \cite{massonis2020finding, structidjl} (for an example in a different modeling context, see \cite{fitzgerald2025practical}). In general, this process can impact the interpretability of state variables of the model or rate parameters. One way to resolve the issue in the case above is to take $\psi=\gamma$, which rescales the parameters $\epsilon$ and the state variable $E$, is enough to break the scaling symmetry and ensure that the model is structurally identifiable if only $C$ can be observed\footnote{This can be verified independently using \texttt{StructuralIdentifiability.jl} \cite{structidjl}}.The reparameterized ``Indirect + Direct'' model would then have the form 
 
 \begin{align*} 
\text{Number of susceptible deer, Unobserved}~\dfrac{dS}{dt} &= a - S(\beta I + \tilde{E} + m), \\
\text{Infected Deer Density, Unobserved}~~\dfrac{dI}{dt} &= S(\beta I + \tilde{E}) - I(m + \mu), \\
\text{Scaled mass of infectious material in the environment, Unobserved}~\dfrac{d\tilde{E}}{dt} &= \tilde{\epsilon} I - \tau  \tilde{E}, \\
\text{Number of dead infected deer, Observed}~\dfrac{dC}{dt} &= \mu I, 
\end{align*} 

where $\tilde{E} = \gamma E$, and $\tilde{\epsilon} = \gamma \epsilon$. 

Of course, whether unique parameter estimates are needed depends on how the model will be used. For example, in the analysis presented in this paper, unique parameter estimates are not needed, as we are concerned with qualitative changes in the system's behavior. However, in different modeling situations (i.e., generating precise forecasts using a model, or evaluating the impact of managerial controls quantitatively), unique parameter estimates may be important, or at least modelers should be conscientious of possible identifiability issues, especially in the context of wildlife management decisions. 
 
We sought to limit the complexity of the model considered here, largely to keep the analytic calculations for the bifurcation analysis tractable (Section \ref{bif}) and allow the model to be easily interpretable. In actuality, some of the model parameters, for example, the carrying capacities, $K$ and $K_w$, may vary over time, as carrying capacities are a dynamic interaction between populations and the environment. We assumed constant parameters to retain simplicity in our model and allow for analysis of the system. Additionally, given the newer findings about the wide range of environmental reservoirs in which prions are found \cite{carlson2023plants,inzalaco2023ticks}, our model may be overly simple. Both ticks and plants have complex, multi-year seasonal life cycles that exhibit phases of dormancy that are cued by environmental signals, potentially inducing a periodic and stochastic component to indirect environmental infection. These complex dynamics are beyond the scope of our model, but may be important to understanding the spread of CWD in deer at the level of an ecosystem as more is understood about the relative importance of different modes of transmission. 

In our analysis, we made several assumptions related to the implementation of the modified Gillespie algorithm and our modeling framework. We took the rate at which susceptible deer, $s$, and wolves, $w$, are born as in Eqs. (\ref{bs_rate}) and (\ref{ws_rate}), respectively. Note that this is a relatively common approximation, but not exact, as it allows for the same ``molecule'' to react with itself. In Eq. (\ref{ws_rate}) we take the rate as $ R_{10}= b_{w} w(1 - {A_0\over K_w' A}w)$ instead of $R_{10} = b_{w} w(1 - {A_0\over K_w' A}(w-1))$ and similarly for Eq. (\ref{bs_rate}). Near the carrying capacity, this is a fine approximation, but less accurate for smaller populations. Relatedly, the population $s$ will always go extinct in finite time for this model and stochastic simulation \cite{childs_keener_keizer}, and so to prevent that from happening, $s$ is artificially never allowed to become zero. Our goal was to maintain stochastically oscillating populations to assess the possibility of controlling CWD outbreaks using wolves. However, it should be noted that the system cannot realize the extinction state when we make this modeling choice, though we would not expect that wolves would hunt a large deer population to extinction. 

In terms of our modeling framework, we chose to work with an ODE model and did not consider the role of possible spatial inhomogeneity. However, the role of space may be important in understanding the spread of CWD in deer. Currently, little is known about the spatial distribution of CWD prions. One would imagine the spatial distribution of the prion is initially somewhat ``patchy'' in space because an infected animal carcass contains a high concentration, which would then diffuse in the local environment through a variety of mechanisms. For instance, it is known that prions can remain infectious after passing through a crow's digestive system \cite{fischer2013could, vercauteren2012prion} and a coyote's digestive system \cite{nichols2015cwd}, which may lead to the spread of $E$ in the environment. The same may be expected of other predators, such as a wolf, and so it may be expected that wolves would spread the prions throughout the environment through feces. However, a different study showed that mountain lions may neutralize CWD prions during digestion \cite{baune2021reduction}. As more is understood about the spread or elimination of the prion by predators, one might consider modifying Eq. (\ref{eq:e}) in our simple model to include a term that describes the spread ($\xi >0$) or elimination ($\xi < 0$) of $E$ by the predator, such as 
 
 $$\frac{dE}{dt} =\epsilon(1 + \xi W)I -\mu_e  E, $$ 
 
 but more biological research is needed to determine how dominant a factor the spread or elimination of the CWD prion by predators may be. Currently, less is known about the transmission of the disease through insects and plants, but these are other mechanisms by which prions can spread throughout the environment in a spatial sense. More research is needed to understand the spatial distribution of prions in the environment and how this contributes to the indirect transmission of CWD. As more is understood about the dynamics of environmental reservoirs of prions, such as ticks and plants, the spatial distribution of prions in the environment, and, more broadly, an ecosystem-scale understanding of CWD transmission, our model can be modified to understand these dynamical processes, or it may be more appropriate to use a spatially explicit model, such as a PDE. 
 
 
This modeling analysis shows that predators may be used as an effective control strategy for managing the spread of CWD in deer populations and highlights a role for hunting or other harvesting of the susceptible population. We are careful to note that for wolves, or other predators, to act as an effective control strategy for CWD outbreaks, strong selective predation of infected deer over susceptible deer is required in our analysis. However, this modeling work should be understood as a proof-of-concept and not used directly for management purposes, given the uncertainty that exists in the parameters of the system. Given the available biological data for this system and associated identifiability challenges, care is needed when using mathematical models for wildlife management purposes. 

 
 
%

\section{Funding}

  C. E. F. is supported in part by The James S. McDonnell Foundation Postdoctoral Fellowship Award in Complex Systems (\url{https://doi.org/10.37717/2020-1591}) and by the NSF-Simons Center for Quantitative Biology at Northwestern University (NSF: 1764421 and Simons Foundation/SFARI 597491-RWC).

\section{Author Contribution}

C.E.F. and J.P.K. designed the research. J.P.K. performed the model analysis and simulations. C.E.F. reviewed the analysis and performed the identifiability analysis. J.P.K. and C.E.F. wrote and edited the paper. 

\section{Code}

The code used for this analysis can be found here \url{https://github.com/cefitzg/cwd_code.git}. ChatGPT-5.5 was used to check the code for bugs, and Grammarly was used to catch typos in the manuscript. 




 \bibliography{./Cervid_Bib}

\end{document}